\documentclass[iop]{emulateapj}
\usepackage{amsmath}
\usepackage{graphicx}
\usepackage{natbib}
\usepackage{epstopdf}
\bibpunct{(}{)}{;}{a}{}{,}
\usepackage{txfonts}
\usepackage{longtable}
\shorttitle{A more centrally concentrated primordial population in  M15}
\shortauthors{S. S. Larsen et al.}
\slugcomment{ApJ, accepted (27 Feb 2015)}

\begin{document}
\title{Radial distributions of sub-populations in the globular cluster M15: a more centrally concentrated primordial population\footnotemark[1]
}
\footnotetext[1]{
Based on observations made with the NASA/ESA Hubble Space Telescope, obtained at the Space Telescope Science Institute, which is operated by the Association of Universities for Research in Astronomy, Inc., under NASA contract NAS 5-26555. These observations are associated with program \#13295.
}


\author{S{\o}ren S. Larsen}
\affil{Department of Astrophysics / IMAPP, Radboud University, PO Box 9010, 6500 GL Nijmegen, The Netherlands}
\email{email: s.larsen@astro.ru.nl}

\author{Holger Baumgardt}
\affil{School of Mathematics and Physics, University of Queensland, St.\ Lucia, QLD 4072, Australia}

\author{Nate Bastian}
\affil{Astrophysics Research Institute, Liverpool John Moores University, 146 Brownlow Hill, Liverpool L3 5RF, United Kingdom}

\author{Jean P. Brodie}
\affil{UCO/Lick Observatory, University of California, Santa Cruz, CA 95064, USA}

\author{Frank Grundahl}
\affil{Stellar Astrophysics Centre, Department of Physics and Astronomy, Aarhus University, Ny Munkegade 120, DK-8000 Aarhus C, Denmark}

\author{Jay Strader}
\affil{Department of Physics and Astronomy, Michigan State University, East Lansing, Michigan 48824, USA}

\begin{abstract}
We examine the radial distributions of stellar populations in the globular cluster (GC) M15, using HST/WFC3 photometry of red giants in the nitrogen-sensitive F343N-F555W color. Surprisingly, we find that giants with ``primordial'' composition (i.e., N abundances similar to those in field stars) are the \emph{most} centrally concentrated within the WFC3 field. We then combine our WFC3 data with SDSS $u,g$ photometry and find that the trend reverses for radii $\ga1\arcmin$ (3 pc) where the ratio of primordial to N-enhanced giants increases outwards, as already found by Lardo et al. The ratio of primordial to enriched stars thus has a U-shaped dependency on radius with a minimum near the half-light radius. N-body simulations show that mass segregation might produce a trend resembling the observed one, but only if the N-enhanced giants are $\sim0.25 \, M_\odot$ less massive than the primordial giants, which requires extreme He enhancement ($Y\ga0.40$). However, such a large difference in $Y$ is incompatible with the negligible optical color differences between primordial and enriched giants which suggest $\Delta Y\la0.03$ and thus a difference in turn-off mass of $\Delta M\la0.04 \, M_\odot$ between the different populations. The radial trends in M15 are thus unlikely to be of dynamical origin and presumably reflect initial conditions, a result that challenges all current GC formation scenarios. We note that population gradients in the central regions of GCs remain poorly investigated and may show a more diverse behavior than hitherto thought.
\end{abstract}

\keywords{globular clusters: individual (M15) --- stars: abundances --- stars: Hertzsprung-Russell and C-M diagrams}

\section{Introduction}

There is strong evidence, both from photometry and spectroscopy, that globular clusters (GCs) have large internal star-to-star variations in the abundances of light elements. While some stars in GCs display the same chemical abundance patterns observed in metal-poor field stars, a large fraction of the GC stars (often the majority) exhibit combinations of light element abundances that are unique to GCs \citep[apart from a small fraction of halo stars that may have escaped from GCs;][]{Martell2011}.
This includes enhanced abundances of He, N, Na, and Al and depleted abundances of C, O, and Mg. Large spreads in the abundances of heavier elements (such as Ca and Fe) are relatively rare, although a significant fraction of the GC population may exhibit small (but detectable) spreads in iron abundance at the level of $\sim0.05$ dex \citep{Carretta2009c,Willman2012}.  

The large and correlated spreads in light element abundances point to proton-capture nucleosynthesis at high temperatures as the main source of the observed abundance anomalies \citep{Cottrell1981,Langer1993}. However, the site where the processing takes place, as well as the mechanism by which processed material is subsequently incorporated into new generations of stars, remain uncertain. The main candidates for the production site are massive asymptotic giant branch (AGB) stars, where the relevant nuclear reactions take place at the bottom of the convective envelope during hot bottom burning \citep{Ventura2001,DAntona2007}, or massive (single or binary) main sequence stars \citep{Wallerstein1987,Brown1993,Prantzos2006,Decressin2007,DeMink2009}. In the AGB scenario, the polluted material is lost from the surface of the stars via slow winds that remain trapped within the gravitational potential of the cluster. In order to explain the observed continuous anticorrelation between [Na/Fe] and [O/Fe], some fraction of the polluted wind material must be diluted with ``pristine'' gas, i.e., gas with the same composition as the original (first-generation) stars \citep{DAntona2007}. 
In the ``winds of fast rotating main sequence stars'' (WFRMS) scenario, it is assumed that massive stars within GCs rotate near break-up speed. Processed material is brought to the surface by rotational mixing, lost via a mechanical wind, and then accumulates in a disk around the star, where the second generation of (low-mass) stars is assumed to form \citep{Decressin2007a,Krause2013}. Alternatively, interacting or merging massive binary stars may provide an efficient way to lose large amounts of processed material that could be incorporated into an enriched population \citep{DeMink2009}. This last scenario is attractive because a large fraction of massive stars are indeed observed to be members of binaries that will interact during their lifetime \citep{Sana2013}. 

In both the AGB and WFRMS scenarios, only a small percentage ($\sim5$\%) of the initial mass of the first generation is returned in the form of polluted material, which leads to a ``mass budget'' problem. The observed large fractions of polluted stars in GCs are accommodated by assuming that most of the first-generation stars have been preferentially lost, implying that GCs were initially a factor of $>10$ more massive than they are now \citep{Decressin2007a,Decressin2010,Vesperini2010}. Such a copious mass loss is, however, difficult to accommodate for the metal-poor GCs in the Fornax, WLM, and IKN dwarf galaxies, among which at least the Fornax GCs show the same anomalies as Galactic GCs \citep{Larsen2014a}. In these dwarf galaxies, we have found that the GCs currently account for at least 20\% of the metal-poor stars, which is difficult to reconcile with a loss of $>90$\% of the initial cluster mass \citep{Larsen2012,Larsen2014}. Other difficulties with these scenarios are that no young massive star clusters with extended, on-going star formation have yet been found \citep[despite extensive searches which have included clusters with masses similar to those expected for young GCs;][]{Bastian2013c,Bastian2014,Cabrera-Ziri2014,Cabrera-Ziri2015}, and that the embedded phase lasts much shorter than expected from the WFRMS scenario \citep{Bastian2014a}.

Alternatively, \citet{Bastian2013a} have suggested that the polluted material lost from interacting binaries may be swept up by accretion disks around low-mass stars. This ``early disk accretion'' scenario might provide a solution to the mass-budget problem, because  (1) large amounts of polluted material are available from the interacting binaries and (2) the material is accreted onto pre-existing low-mass stars. In this scenario, there is thus only a single ``generation'' of stars. The time scale for the accretion to take place may, however, require some fine tuning to ensure that a sufficient amount of ejecta are accreted and mixed while the low-mass stars are still in the convective phase \citep{DAntona2014}.
In the remainder of this paper, we will generally refer to the stars with field-like composition as ``primordial'' and those that have modified light-element abundances as ``enriched'' when discussing our observations, and thereby avoid implying a particular sequence of events.

All of the above scenarios predict that the enriched stars should be located preferentially in the central regions of the clusters. In the AGB scenario, the first (primordial) generation is envisioned to expand following expulsion of gas left-over from the initial burst of star formation, after which wind material accumulates in the center via a cooling flow and forms a more centrally concentrated enriched population \citep{DErcole2008}. In the WFRMS scenario, the enriched stars are also expected to form preferentially in the central regions, because they form in the vicinity of mass-segregated massive stars \citep{Decressin2008}.  A similar prediction is made by the early disk accretion model because accretion is more efficient in the central regions where the density is higher. These expectations appear to be borne out by observations of some GCs, where a number of studies have found the enriched stars to be distributed preferentially near the center \citep{Norris1979,Carretta2009a,Kravtsov2010,Lardo2010,Milone2012a}. However, the differences in spatial distribution are expected to be eventually erased by dynamical evolution. This should happen first in the central regions of clusters, where the relaxation time is shortest \citep{Vesperini2013}. Indeed, it has recently been found that the two populations in the cluster NGC~6362 do \emph{not} exhibit any differences in their radial distributions \citep{Dalessandro2014}.
 
To a large extent, then, the spatial distributions of stellar populations within most GCs observed to date appear consistent with theoretical expectations that are common to all formation scenarios. However, there may be more subtle differences between the predictions of different scenarios. One such difference concerns stars with intermediate composition, which are supposed to have formed out of diluted wind material in the AGB model and would therefore have been the \emph{last}  to have formed. In the early disk accretion scenario, the intermediate population would instead correspond to stars that did not pass through the densest part of the cluster. It is, therefore, a clear prediction of this scenario that such stars should have an intermediate degree of central concentration. In the other scenarios it is less clear what to expect, but it seems plausible that the intermediate-composition population may be expected to be the \emph{most} centrally concentrated in the AGB scenario, since it is the last to form. While this remains somewhat speculative, it does suggest that interesting constraints on formation scenarios may be obtained by studying the radial distributions in more detail.

In light of the scenarios outlined above, we have used Hubble Space Telescope (HST) observations to examine the spatial distributions of red giants in the globular cluster M15 (NGC~7078) as a function of their chemical composition. 
With a metallicity of $\mathrm{[Fe/H]}=-2.3$ \citep{Carretta2009c}, M15 is one of the most metal-poor GCs in the Milky Way. While the internal spread in $\mathrm{[Fe/H]}$ is small, estimated at $\sigma_\mathrm{[Fe/H]}\sim0.05$ dex \citep{Carretta2009c,Willman2012}, M15 is similar to other Galactic GCs in showing large internal abundance variations of the light elements. Observations of red giants have revealed the well-known Na/O anti-correlation, as well as a clear Mg/Al anti-correlation \citep{Sneden1997}. The abundances of C and N also exhibit large and anti-correlated variations with [C/Fe] and [N/Fe] ratios varying by $\sim1$ dex and  $\sim2$ dex, respectively \citep{Trefzger1983,Cohen2005a,Pancino2010a}. In addition to the light-element abundance variations, M15 is one of a few GCs that are known to exhibit a spread in the abundances of the heavy ($n$-capture) elements Sr, Y, Zr, Ba, La, and Eu, which however do not appear to correlate with the  abundance variations of the light elements and may be of a different origin \citep{Sneden1997,Otsuki2006,Sobeck2011}. 
In spite of relatively short exposure times, our data are sensitive to N abundance variations for stars at the base of the red giant branch (RGB). Combined with the high luminosity of the cluster \citep[$M_V\sim-9.1$;][]{Harris1996}, this yields a sample of more than 1300 red giants, which allows us to examine radial trends of the sub-populations in some detail.  
In this paper, we report the (unexpected) results of our investigation.

\section{Observations and data reduction}

As part of our HST program to study the globular clusters in the Fornax dwarf spheroidal galaxy \citep[][hereafter Paper I]{Larsen2014a}, we obtained short exposures of M15 in the same filters as those used for our main program: F343N, F555W, and F814W  (Program ID: GO-13295, P.I.: S. S. Larsen). These observations exploit the well-established sensitivity of ultraviolet photometry to light element abundance variations \citep{Hesser1977,Grundahl2002,Yong2008,Sbordone2011,Monelli2013}.
The integration time was $2\times350$ s in F343N and $2\times10$ s in F555W and F814W with M15 centered on CCD \#2 of the Wide Field Camera 3 (WFC3). The post-flash option was used to mitigate the effect of charge transfer losses by increasing the background level to 10 counts per pixel. All exposures were obtained within a single orbit. Note that these exposures were not designed to be very deep, but were only intended to reach stars on the red giant branch. 

The pipeline-reduced images were corrected for charge transfer inefficiencies with the program \texttt{wfc3uv\_ctereverse}\footnote{\texttt{http://www.stsci.edu/hst/wfc3/tools/cte\_tools}}. We then used the \texttt{astrodrizzle} code to align, combine, and resample the two exposures in each filter to a uniform pixel scale of $0\farcs040$ per pixel.
Point-spread function (PSF) fitting photometry was carried out with \texttt{ALLFRAME} \citep{Stetson1994} and calibrated to standard STMAG magnitudes as described in Paper~I.

In addition to our own WFC3 data, we use imaging of M15 in the F606W and F814W filters obtained with the Advanced Camera for Surveys (ACS) on HST as part of the ACS Galactic Globular Cluster Survey \citep[ACSGCS;][]{Sarajedini2007}. The ACSGCS data consist of short exposures with integration times comparable to those of our F555W/F814W data (15 s in each filter), as well as deeper exposures ($4\times130$ s in F606W, $4\times150$ s in F814W) that allow accurate photometry for stars well below the main sequence turn-off. We did not carry out photometry on these images ourselves, but use the catalogs published by the ACSGCS team \citep{Anderson2008}.

Throughout this paper we assume a distance of 10.3 kpc \citep{VandenBosch2006}, along with a foreground extinction of $A_V=0.30$ mag \citep{Schlafly2011}. Using the \citet{Cardelli1989} extinction law, this yields $A_{F343N} = 0.483$ mag, $A_{F555W} = 0.312$ mag, $A_{F606W} = 0.276$ mag, and $A_{F814W} = 0.178$ mag in the HST filters.

\section{Results}

\subsection{Photometric evidence for a spread in the N abundance}
\label{sec:nvar}

\begin{figure}
\includegraphics[width=\columnwidth]{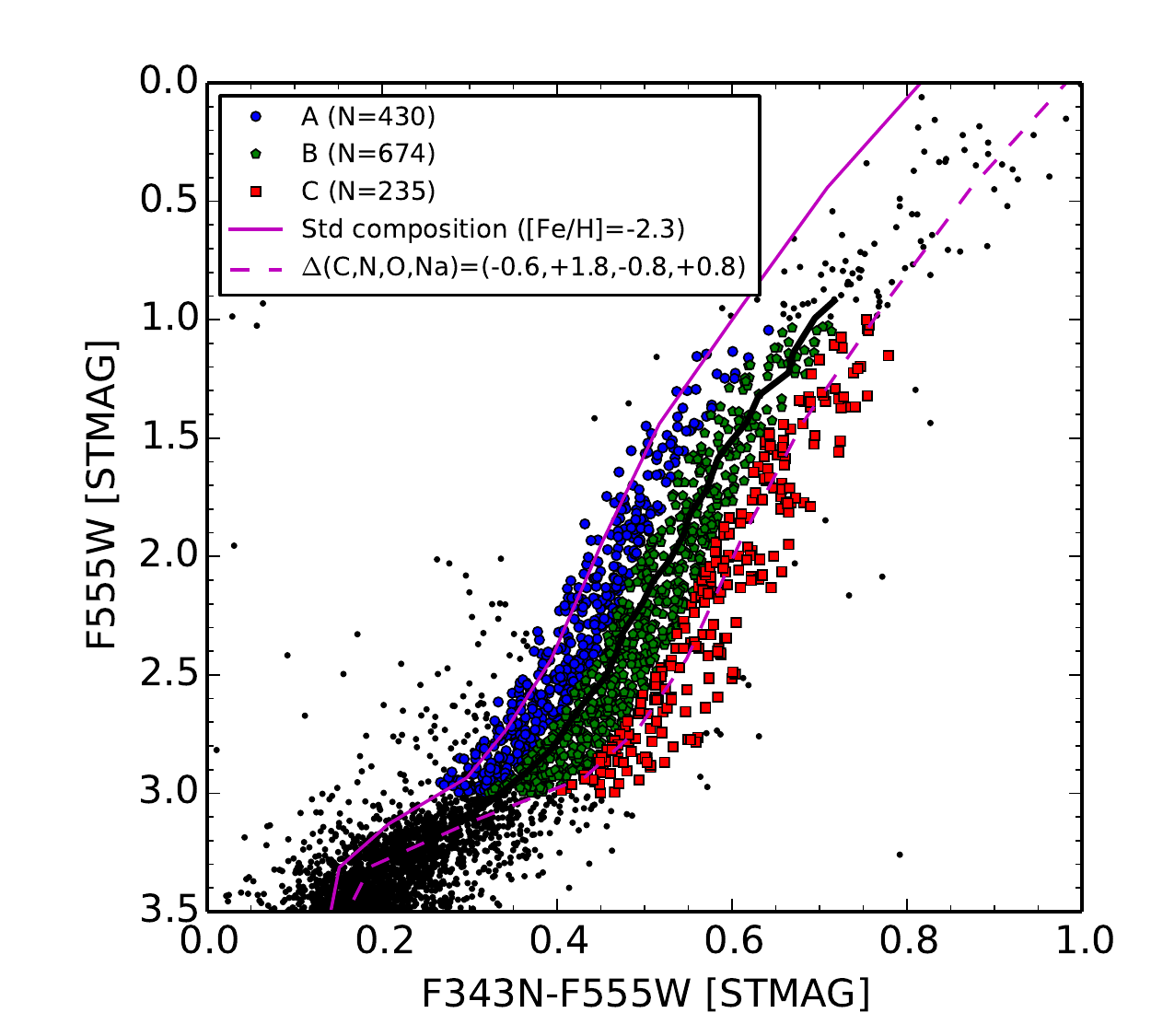}
\caption{\label{fig:m15sel}(F343N-F555W,F555W) color-magnitude diagram showing the lower RGB of M15. Symbols are color-coded according to the division into normal, intermediate, and strongly enhanced N abundances (group A, B, and C). The thick black line is a median ridge line.}
\end{figure}

\begin{deluxetable*}{lccccccccccc}
\tablecaption{\label{tab:phot}ALLFRAME photometry of M15}
\tablecolumns{12}
\tablehead{
 ID  & X & Y & R.A. & Decl. & R & \multicolumn{2}{c}{F343N} & \multicolumn{2}{c}{F555W} & \multicolumn{2}{c}{F814W} \\
 & & & & & & \colhead{mag}   & \colhead{err} 
 & \colhead{mag}   & \colhead{err} 
 & \colhead{mag}   & \colhead{err} 
}
\tabletypesize{\footnotesize}
\startdata
   102 &  658.22 &  51.70 & 322.508246 & 12.154236 & 70.86 & 18.607 & 0.036 & 17.975 & 0.017 & 18.345 & 0.018 \\
   108 &  534.34 &  54.74 & 322.509641 & 12.154211 & 74.80 & 18.874 & 0.037 & 18.318 & 0.020 & 18.688 & 0.019 \\
   272 &  304.00 &  84.04 & 322.512246 & 12.154424 & 81.94 & 19.176 & 0.043 & 18.904 & 0.031 & 19.499 & 0.026 \\
   338 & 1138.58 &  91.72 & 322.502863 & 12.154904 & 55.56 & 18.820 & 0.029 & 18.430 & 0.012 & 18.887 & 0.010 \\
   353 &  952.61 &  93.77 & 322.504955 & 12.154838 & 60.74 & 18.518 & 0.021 & 17.922 & 0.020 & 18.274 & 0.020 \\
   355 & 1363.93 &  93.60 & 322.500329 & 12.155031 & 49.90 & 18.995 & 0.016 & 18.647 & 0.020 & 19.138 & 0.019 \\
   374 & 1236.21 &  95.60 & 322.501766 & 12.154993 & 52.90 & 19.285 & 0.027 & 18.981 & 0.014 & 19.604 & 0.020 \\
   389 & 1194.67 &  97.66 & 322.502235 & 12.154996 & 53.90 & 19.337 & 0.025 & 18.964 & 0.011 & 19.632 & 0.018 \\
   406 & 1160.12 & 100.44 & 322.502624 & 12.155010 & 54.72 & 19.111 & 0.027 & 18.748 & 0.013 & 19.263 & 0.018 \\
   463 & 1605.34 & 105.80 & 322.497620 & 12.155280 & 44.70 & 17.613 & 0.023 & 16.942 & 0.016 & 17.296 & 0.012
\enddata
\tablecomments{X and Y are the coordinates in the drizzled CCD frames. R is the projected distance from the cluster center (in arcsec). For each filter, the magnitude and error estimate by \texttt{ALLFRAME} are listed.
(This table is available in its entirety in a machine-readable form in the online journal. A portion is
shown here for guidance regarding its form and content).}
\end{deluxetable*}

Figure~\ref{fig:m15sel} shows the (F343N-F555W, F555W) color-magnitude diagram for the lower part of the M15 RGB  (a larger color/magnitude range, as well as the (F555W-F814W, F555W) CMD, are shown in Paper I). 
Here, and in the following, we only include stars in the radial range $4\arcsec<R<130\arcsec$ for which the \texttt{ALLFRAME} photometry has $\chi^2_\nu < 3$ in the F555W images. Closer to the center, crowding prevents accurate photometry and for $R>130\arcsec$ there are very few stars although the outermost corner of the WFC3 mosaic is nominally located at $R\sim150\arcsec$ from the center of M15. Table~\ref{tab:phot} lists the photometry for all stars brighter than $\mathrm{F555W}=19$ ($M_\mathrm{F555W} = 3.6$) that meet these criteria.

As in Paper~I, we will generally exclude stars brighter than $M_\mathrm{F555W}=1$ from our analysis, because their surface abundances may have been modified by deep mixing \citep{Gratton2000}. However, because of the better S/N of the M15 data, we can obtain good photometry for somewhat fainter RGB stars $(1.0 < M_\mathrm{F555W} < 3.0)$ than in Paper I, where we adopted a limit of $M_\mathrm{F555W}=2.5$ mag. The symbols for stars in this magnitude range are color-coded in Fig.~\ref{fig:m15sel} according to their offset in the (F343N-F555W) direction, as discussed further below (Sec.~\ref{sec:radial}). On first inspection, we note that the spread of the RGB stars in F343N-F555W is far greater than the photometric uncertainties, which are $\sim0.02$ mag (Sec.~\ref{sec:artstar}). Furthermore, the spread does not depend significantly on the location along the RGB, again consistent with most of it being real.

We also plot model colors for N-normal composition and the N-enhanced ``CNONaI'' mixture \citep{Sbordone2011} which has $\Delta$([C/Fe], [N/Fe], [O/Fe], [Na/Fe]) = $(-0.6, +1.8, -0.8, +0.8)$ dex relative to standard ($\alpha$-enhanced) composition. The colors were computed for a 13 Gyr isochrone \citep{Dotter2007} with $\mathrm{[Fe/H]}=-2.3$ and $[\alpha/\mathrm{Fe}]=+0.4$ by integrating \texttt{ATLAS12}/\texttt{SYNTHE} model spectra \citep{Sbordone2004,Kurucz2005} over the filter transmission curves. The two model lines are clearly separated, with the N-enhanced models being redder by about 0.16 mag, and the observed F343N-F555W colors span a range comparable to, or somewhat greater than, the separation between the models. This is in agreement with the spectroscopically measured N abundance spread in M15 of about 2 dex \citep{Cohen2005a}. 
We note that the models do not match the sub-giant branch perfectly but appear slightly too bright/blue. A better match to the subgiants can be achieved by increasing the extinction correction by $\Delta A_V \sim 0.02$ mag and the assumed distance by $\sim$0.25 kpc, which is well within the 0.4 kpc uncertainty \citep{VandenBosch2006}. However, no conclusions in this paper are affected by these small adjustments and in what follows we simply keep the literature values.

\begin{figure}
\includegraphics[width=80mm]{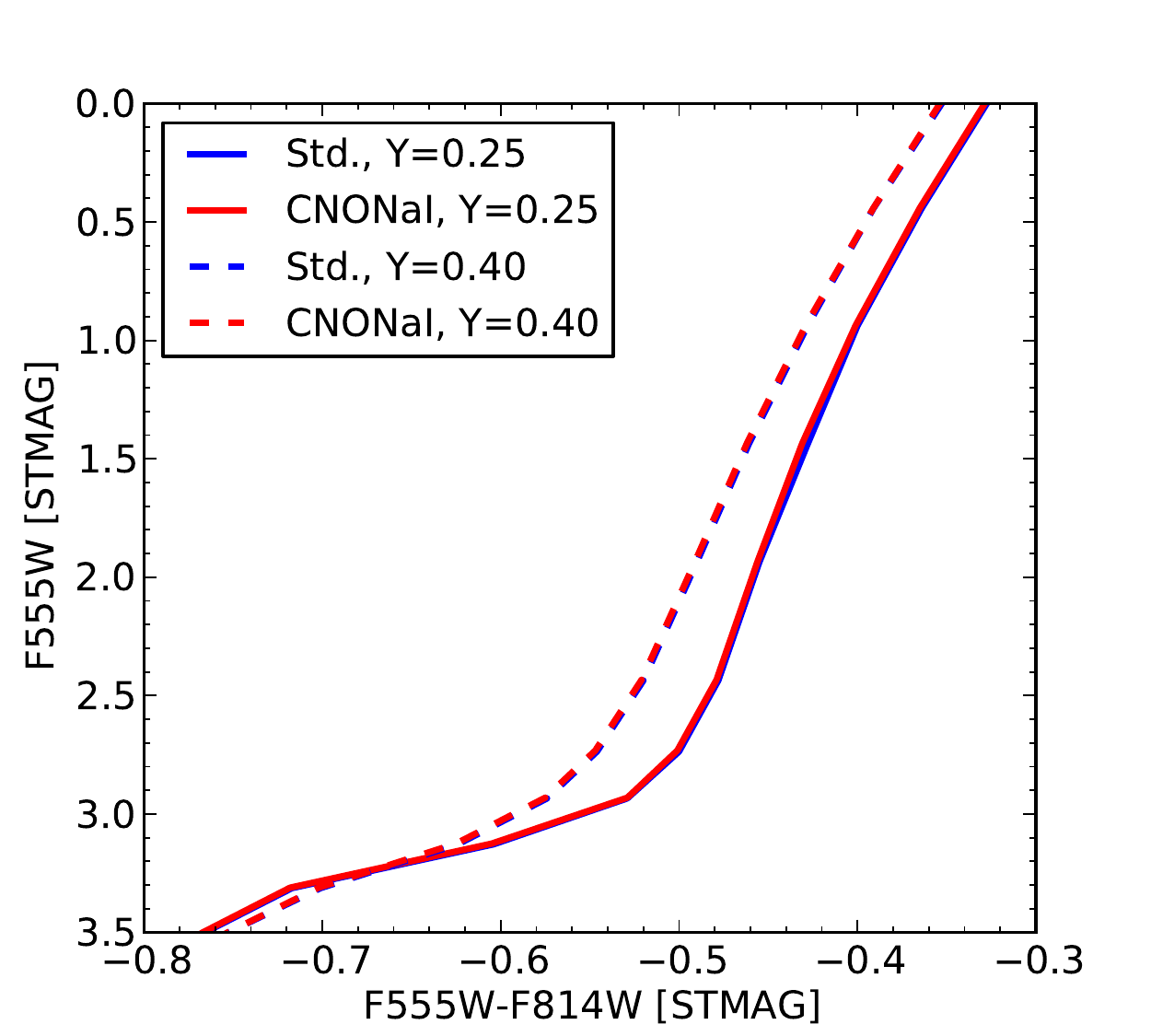}
\includegraphics[width=80mm]{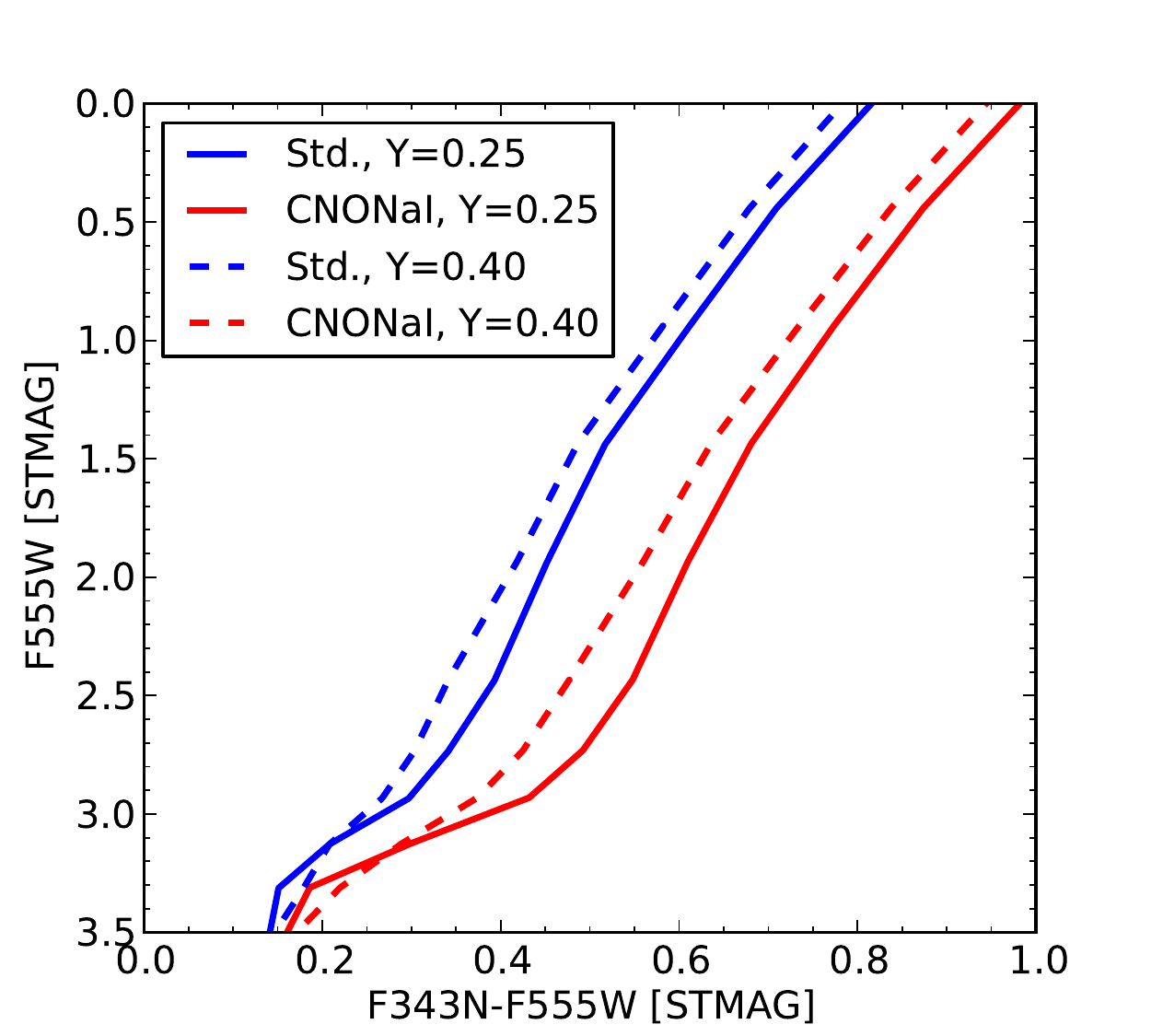}
\caption{\label{fig:colcmp}Model colors for different He and C,N,O,Na abundances. The isochrones have an age of 13 Gyr and $\mathrm{[Fe/H]}=-2.3$. The F555W-F814W colors are essentially independent of the CNONa mixture, but sensitive to He abundance, whereas the F343N-F555W colors are mainly sensitive to CNONa variations.}
\end{figure}

Apart from the color variations in F343N-F555W that arise from variations in the light element mixture, the color-magnitude diagram can also be affected by He abundance variations  \citep{Norris2004,Salaris2006}. While the variations in F343N-F555W color are mainly an atmospheric effect, due to the molecular absorption bands in the UV (predominantly the NH band near 3370 \AA), He abundance variations modify the internal structure and effective temperature of the stars and therefore also affect optical colors. These effects are illustrated in Fig.~\ref{fig:colcmp} for the WFC3 filters used in our program. We have combined isochrones for $Y=0.25$ and $Y=0.40$ with \texttt{ATLAS12/SYNTHE} model atmospheres and synthetic spectra computed specifically for these $Y$ values and normal and CNONaI light-element mixture. A fixed iron abundance, relative to hydrogen, of $\mathrm{[Fe/H]}=-2.3$ was assumed in all cases. As found by other authors, the $\mathrm{F555W}-\mathrm{F814W}$ colors of stars on the lower RGB are virtually insensitive to the CNONa abundances, but become bluer for increased He abundance (upper panel). In contrast, the $\mathrm{F343N}-\mathrm{F555W}$ colors are very sensitive to the CNONa abundances and become redder as the N abundance increases (lower panel). This shift is much greater than the shift toward blue colors  caused by a high He fraction. We thus expect that a N-enriched population will indeed have redder $\mathrm{F343N}-\mathrm{F555W}$ colors than a population with primordial composition, \emph{even} if the N-enriched population is also strongly He enhanced.

\subsection{Differential reddening}
\label{sec:dred}

In addition to the mean foreground reddening toward M15, reddening variations on smaller scales may be present within the field of view of the HST cameras and could potentially affect analyses of spatial trends. \citet{Milone2012b} used the mean colors of main sequence stars to map the reddening  across GCs and found significant variations across the HST/ACS field of view in regions of high foreground reddening ($E(B-V) > 0.1$). For GCs in regions of lower foreground reddening, small reddening variations are difficult to disentangle from systematic variations in the photometric zero-points across the field that may be caused by uncertainties in the PSF modeling. Indeed, \citet{Anderson2008} found systematic variations of about $\pm0.01$ mag in the mean F606W-F814W colors of main sequence stars even for GCs in regions of very low foreground reddening. Since the reddening toward M15 is not entirely negligible, it is worth examining whether there is evidence of differential reddening in the HST data.

\begin{figure}
\includegraphics[width=\columnwidth]{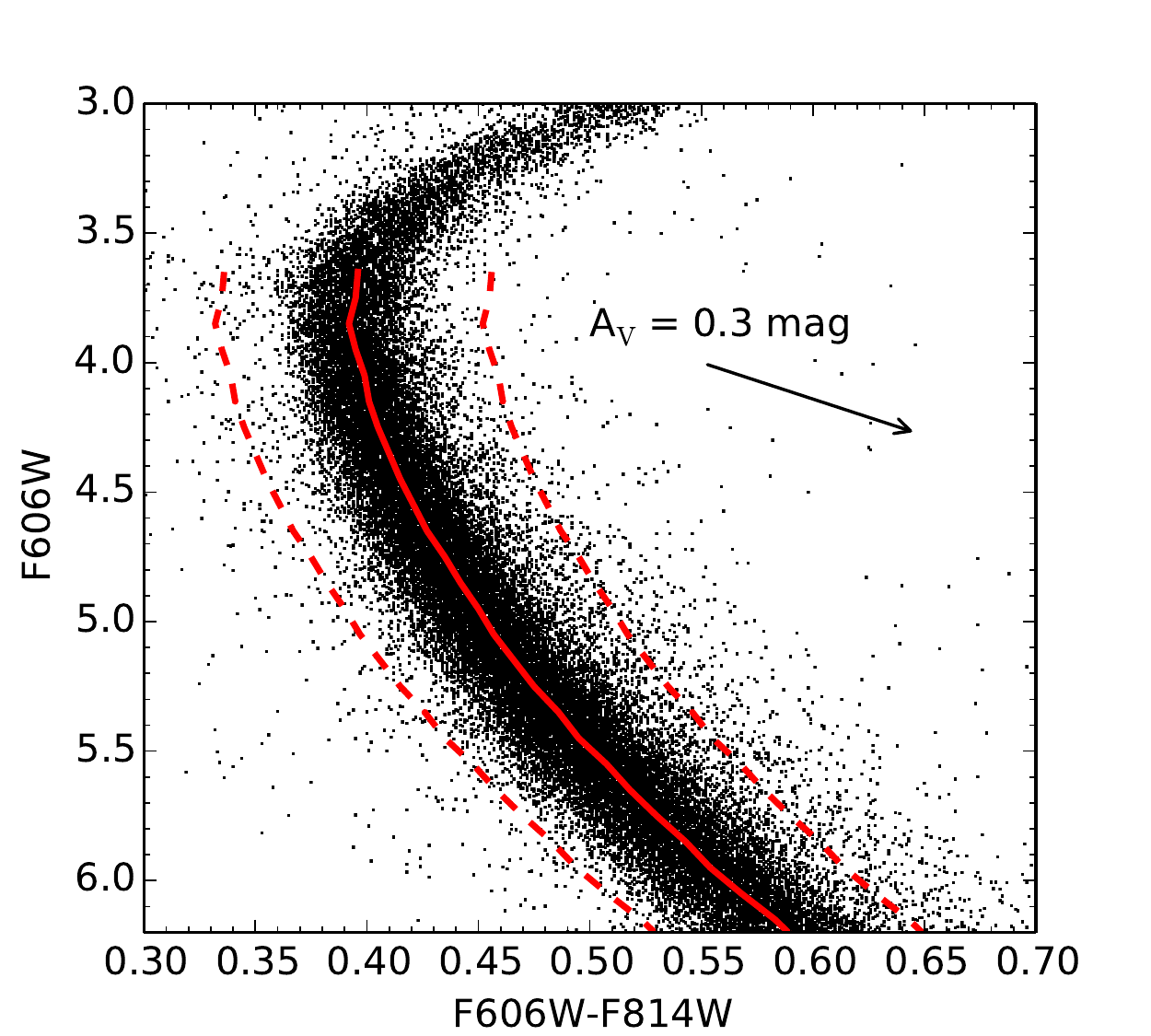}
\caption{\label{fig:ms_io}(F606W-F814W,F606W) CMD for the upper main sequence of M15. The ridge line and the limits used for selection of stars for the reddening map are shown. The arrow is the reddening vector for $A_V=0.3$ mag.}
\end{figure}

Our WFC3 data are not deep enough to allow reliable measurements of main sequence stars, so we used the ACSGCS photometry to map variations in the colors of main sequence stars.  A small fraction of the WFC3 field falls outside the area covered by the ACSGCS data, but this only affects regions with a projected distance of $>140\arcsec$ from the center of M15. We first defined a ridge line by computing the median $\mathrm{F606W}-\mathrm{F814W}$ colors of main sequence stars as a function of F606W magnitude in bins of 0.1 mag. We only included stars with a photometric error in F606W of less than 0.02 mag and with a ``quality-of-fit'' parameter \citep{Anderson2008} of $\mathrm{qfitV}<0.3$. Figure~\ref{fig:ms_io} shows the upper part of the main sequence with the ridge line overplotted.

We then calculated the offset $\Delta_{606-814}$ in the $\mathrm{F606W}-\mathrm{F814W}$ color with respect to the ridge line for each star in the range $4.0 < M_\mathrm{F606W} < 6.0$. Stars with $\left| \Delta_{606-814}\right| > 0.06$  mag were excluded (dashed curves in Fig.~\ref{fig:ms_io}), leaving a total of 29515 stars. The dispersion of these stars around the ridge line was $\sigma = 0.020$ mag in the $\mathrm{F606W}-\mathrm{F814W}$ color. 
The $\Delta_{606-814}$ offsets were converted to estimates of the reddening by taking into account the slope of the ridge line and the reddening vector, 
\begin{equation}
\begin{split}
  E(\mathrm{F606W}-\mathrm{F814W}) = \\ \Delta_{606-814} \left[1 - \left(\frac{\delta_{606-814}}{\delta_{606}}\right) \left(\frac{A_{F606W}}{A_{F606W}-A_{F814W}}\right)\right]^{-1}
  \label{eq:evi}
\end{split}
\end{equation}
where $\frac{\delta_{606-814}}{\delta_{606}}$ is the (inverse) slope of the ridge line, evaluated locally at the magnitude of each star. This relation is valid for offsets that are small enough so that the curvature of the ridge line can be neglected locally. For any individual star, the offset from the ridge line is typically dominated by random photometric errors, but by averaging the offsets of many stars we could map systematic variations in the mean color across the field. 

The sky coordinates from the \citet{Anderson2008} catalog were transformed to pixel coordinates in the WFC3 frame with the \texttt{rd2xy} task in the \texttt{drizzlepac} package. 
At each pixel in the WFC3 image, we computed a weighted average of the reddening values of the surrounding stars, with weights given by a Gaussian function of the distance, $w_i = \exp\left(-\frac{1}{2} \frac{r_i^2}{(100 \, \mathrm{pixels})^2} \right)$ for a star located $r_i$ pixels from a given point
\citep{Larsen1996}. On average, this yielded about 50 stars per ``resolution element'' of the resulting map, although the stellar density obviously varies greatly across the field. 

\begin{figure}
\includegraphics[width=\columnwidth]{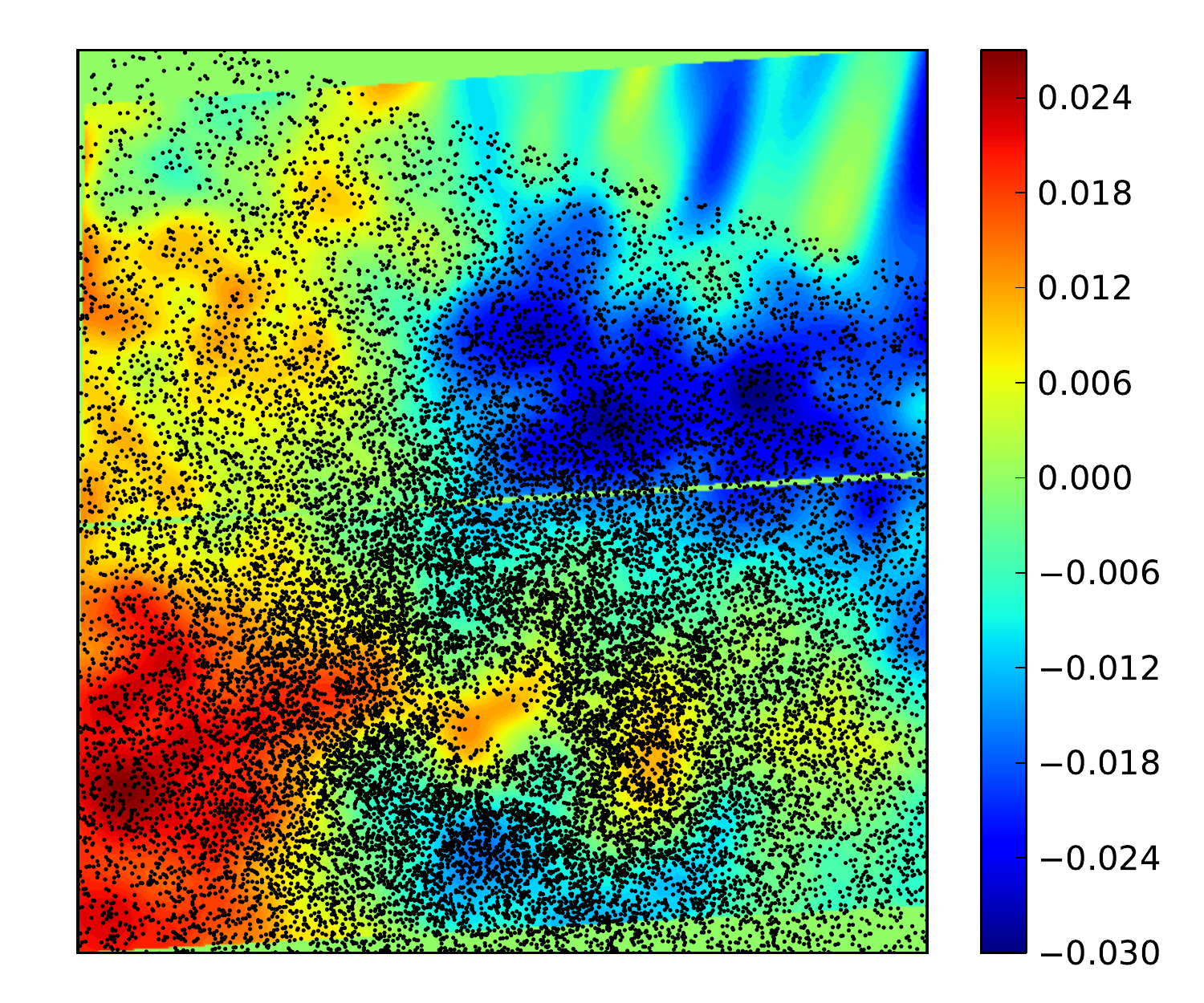}
\caption{\label{fig:extmap}$E(\mathrm{F606W}-\mathrm{F814W})$ reddening map based on $\mathrm{F606W-F814W}$ colors of main sequence stars.  The r.m.s.\ variation in $E(\mathrm{F606W-F814W})$ across the field is $\sigma=0.011$ mag, with peak-to-peak variations of $\sim0.05$ mag. 
}
\end{figure}

The map is shown in Fig.~\ref{fig:extmap}. The color scale indicates the average computed reddening $\langle E(\mathrm{F606W}-\mathrm{F814W})\rangle$ at each position and the black points show the locations of the stars used to produce the map. The gap in the distribution of stars near the center of the lower CCD detector corresponds to the center of M15, and it can be seen that the upper right-hand corner of the WFC3 field falls outside the coverage of the ACSGCS data. Note that the computed color may vary even in regions with few (or no) stars, depending on how the relative weights of the closest stars to any given point change with position. This is seen in the upper right-hand corner, as well as in the region near the center. It is clear that the significance of any features in these empty regions is low.

By comparing maps from two independent sub-samples, we found that the large-scale features in Fig.~\ref{fig:extmap} were consistently reproduced: there is an overall gradient across the field with redder colors (that may be interpreted as higher reddening) in the bottom left part of the field and bluer colors in the top right-hand part. There is some evidence for a filamentary structure extending across the lower half of the field from left to right, whereas most structure on smaller scales appears not to be significant. The r.m.s.\ deviation in $E(\mathrm{F606W}-\mathrm{F814W})$ across the map is 0.011 mag, which is comparable to the variations expected from uncertainties in the PSF modeling \citep{Anderson2008,Milone2012b}.

Apart from differential reddening and position-dependent uncertainties in the photometric calibration, other factors that might affect the optical colors include He abundance, overall metallicity, and binarity. If variations in these quantities were significant, we would expect them to show up as a radial gradient with respect to the center of M15. Given that the structure in Fig.~\ref{fig:extmap} is not obviously symmetric around M15, it seems unlikely that these factors contribute significantly to the structure in the map.
However, it remains difficult to disentangle reddening variations from variations in the photometric zero-points.  The dispersion of the MS stars in $\Delta_{606-814}$ decreased from $\sigma=0.020$ mag to $\sigma = 0.017$ mag when correcting the photometry for reddening according to the map, which is consistent with the 0.011 mag dispersion in $E(\mathrm{F606W}-\mathrm{F814W})$ across the field. However, this decrease in the dispersion of the corrected colors is expected whether the variations are caused by differential reddening or photometric zero-point variations.
A decrease was also seen for the dispersion across the RGB; without any correction the dispersion was $\sigma=0.020$ mag in F606W-F814W, and when applying the reddening map this decreased (slightly) to $\sigma=0.019$ mag. However, in the WFC3 data, the dispersion of the RGB stars in F555W-F814W actually \emph{increased} (from $\sigma=0.024$ mag to $\sigma=0.026$ mag) if a reddening correction was applied. This suggests that the color variations in Fig.~\ref{fig:extmap} are primarily caused by instrumental effects (and thus do not reproduce between the ACS and WFC3 observations), rather than by differential reddening.

Thus having considered the possible effect of differential reddening carefully, we have at the end chosen \emph{not} to correct for it in our general analysis. However, when relevant we will comment on any differences that arise from including or omitting this correction.

\subsection{Artificial star tests}
\label{sec:artstar}

To quantify the photometric errors in the WFC3 observations, we carried out artificial star experiments following the general procedure described in Paper~I. We started by generating coordinate lists for a number of concentric annuli around M15. The annuli covered the radial intervals 100--200 pixels, 200--400 pixels, 400--600 pixels, and 800--1000 pixels. In each annulus, 500 pseudo-random star coordinates were generated by arranging the stars in a polar grid with a spacing of 20 pixels in the radial direction and a spacing in the azimuthal direction that provided the desired total number of stars. A further random dither offset in the range $-0.5\ldots+0.5$ pixels was added to each coordinate in the $x$- and $y$-directions. The area of the innermost bin was too small to accommodate 500 stars with a minimum separation of 20 pixels, so for this bin only 100 coordinates were defined. We then generated lists of F343N, F555W, and F814W magnitudes for the artificial stars by selecting the F555W magnitudes of the actual RGB stars in M15 and interpolating in the N-normal model isochrone to find the other magnitudes. The artificial stars were added to the images using the \texttt{mksynth} task in \texttt{baolab} \citep{Larsen1999}, including a set of artificial PSF stars. The \texttt{ALLFRAME} photometry procedure was repeated, and the artificial stars were recovered by requiring a match within a distance of 1 pixel from the input coordinates. This procedure was repeated four times.

\begin{figure}
\includegraphics[width=\columnwidth]{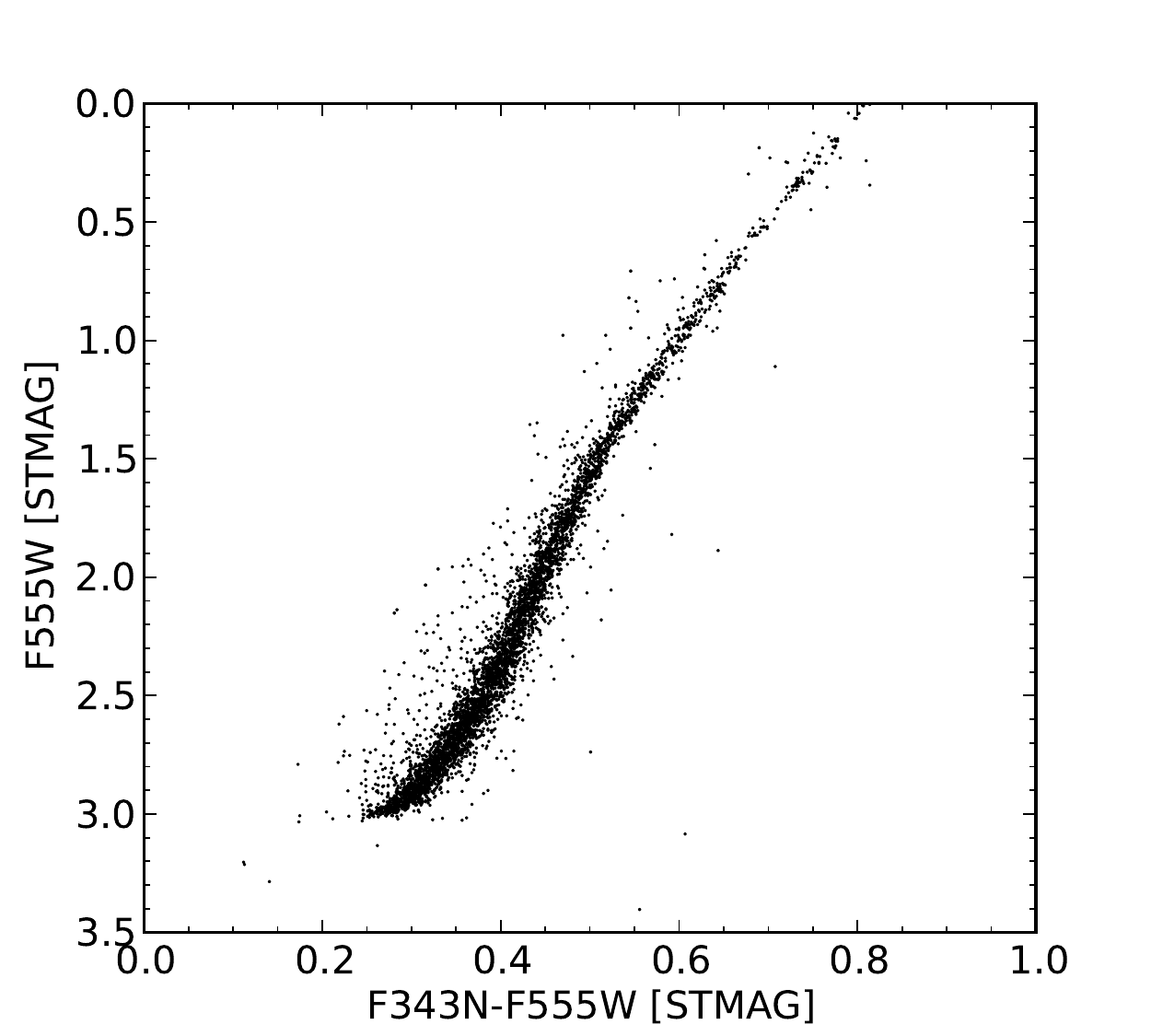}
\caption{\label{fig:syncmd}(F343N-F555W,F555W) CMD for synthetic stars populating a single isochrone. The broadening of the RGB due to photometric errors is far smaller than the observed width of the RGB seen in Fig.~\ref{fig:m15sel}.}
\end{figure}

\begin{figure}
\includegraphics[width=\columnwidth]{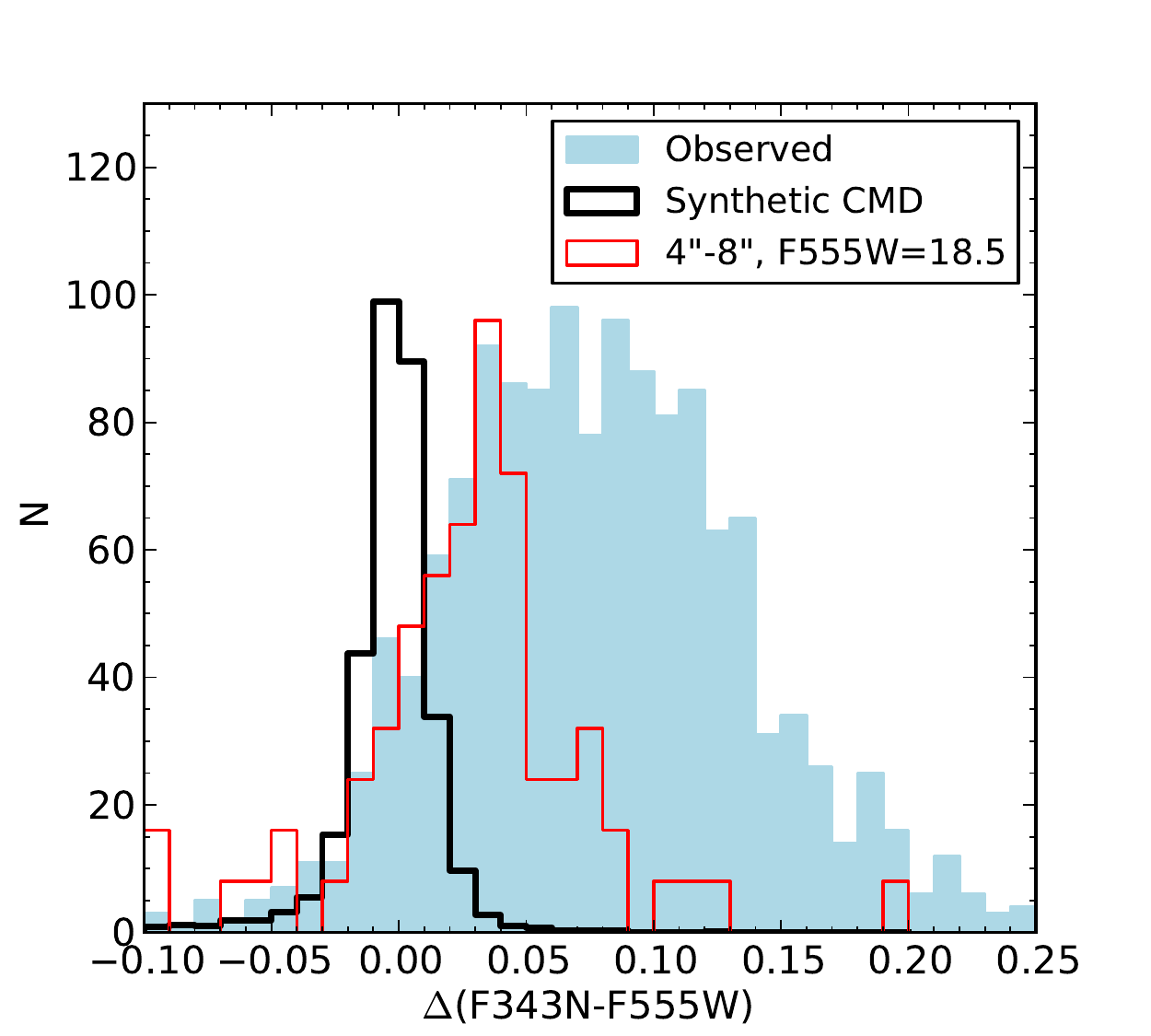}
\caption{\label{fig:uvcmp}Distributions of $\Delta$(F343N-F555W) values for red giants in M15 (filled histogram) and the synthetic CMD (unfilled black histogram). The thin red histogram shows the F343N-F555W distribution for artificial stars with $\mathrm{F555W}=18.5$ mag in the innermost annulus, $R=4\arcsec-8\arcsec$ (cf.\ Table~\ref{tab:photerr}). The red histogram has been shifted by 0.03 mag.}
\end{figure}

Figure~\ref{fig:syncmd} shows the CMD of the recovered artificial stars. By comparison with Fig.~\ref{fig:m15sel}, it is clear that the dispersion in the artificial CMD is much smaller than the observed spread in the F343N-F555W colors. 
As in Paper I, we defined the $\Delta$(F343N-F555W) parameter as the offset between the isochrone of N-normal composition and the observed F343N-F555W color at a given F555W magnitude. In Fig.~\ref{fig:uvcmp} we compare  the distributions of $\Delta$(F343N-F555W) for the observations of red giants in M15 with the artificial star tests for the magnitude range $1 < M_\mathrm{F555W} < 3$. Even though the radial density distribution of the artificial stars is not fully realistic, the histogram for the artificial stars is much narrower than for the observed RGB stars. Formally, the dispersion of the observed $\Delta$(F343N-F555W) values is 0.059 mag, whereas the corresponding dispersion for the artificial stars is 0.018 mag. In $\Delta(\mathrm{F555W}-\mathrm{F814W})$, the synthetic CMD has $\sigma=0.020$ mag, which is slightly less than the observed spread for the RGB stars ($\sigma\sim0.024$ mag). This is consistent with an additional $\sim0.01$ mag variation from uncertainties in PSF modeling and/or differential reddening.

\begin{deluxetable*}{lccccccccc}
\tablecaption{\label{tab:photerr}Photometric errors and completeness from artificial star tests}
\tablecolumns{10}
\tablehead{
 Radius  & \multicolumn{3}{c}{$\mathrm{F555W}=17$} & \multicolumn{3}{c}{$\mathrm{F555W}=18$} & \multicolumn{3}{c}{$\mathrm{F555W}=18.5$} \\
 & \colhead{$\sigma_\mathrm{V-I}$}   & \colhead{$\sigma_\mathrm{U-V}$} & $f_\mathrm{rec}$ 
 & \colhead{$\sigma_\mathrm{V-I}$}   & \colhead{$\sigma_\mathrm{U-V}$} & $f_\mathrm{rec}$
 & \colhead{$\sigma_\mathrm{V-I}$}   & \colhead{$\sigma_\mathrm{U-V}$} & $f_\mathrm{rec}$
}
\tabletypesize{\small}
\startdata
4\arcsec--8\arcsec & 0.013 & 0.013 & 0.99 & 0.029 & 0.032 & 0.99 & 0.045 & 0.038 & 0.99 \\
8\arcsec--16\arcsec & 0.011 & 0.013 & 0.99 & 0.021 & 0.024 & 0.97 & 0.025 & 0.031 & 0.98 \\
16\arcsec--24\arcsec  & 0.011 & 0.011 & 0.99 & 0.017 & 0.018 & 0.99 & 0.021 & 0.022 & 0.99 \\
32\arcsec--40\arcsec & 0.009 & 0.009 & 1.00 & 0.016 & 0.015 & 0.99 & 0.019 & 0.017 & 1.00
\enddata
\tablecomments{$\sigma_\mathrm{V-I}$ and $\sigma_\mathrm{U-V}$ denote the standard deviation of the recovered artificial star colors in F555W-F814W and F343N-F555W, respectively. $f_\mathrm{rec}$ is the fraction of the input artificial stars recovered by the photometry procedure.}
\end{deluxetable*}

To quantify the dependency of the photometric errors on magnitude and radial position further, we carried out a second set of artificial star tests in which stars with fixed magnitudes of
$\mathrm{F555W}=17$, 18, and 18.5 ($M_\mathrm{F555W}\approx1.6$, 2.6, and 3.1) were added to the images. The radial bins and numbers of stars in each bin were the same as previously described.
The resulting dispersions in $\mathrm{F343N}-\mathrm{F555W}$ ($\sigma_\mathrm{U-V}$) and in $\mathrm{F555W}-\mathrm{F814W}$ ($\sigma_\mathrm{V-I}$) are listed in Table~\ref{tab:photerr} for each radial bin and each input magnitude. As expected, crowding causes the errors to increase toward the center, especially for radii $<200$ pixels (8\arcsec), but the errors remain much smaller than the observed color spread at all radii and magnitudes. We have included the recovered F343N-F555W distribution for the innermost annulus ($4\arcsec-8\arcsec$) and $\mathrm{F555W}=18.5$ as a thin (red) histogram in Fig.~\ref{fig:uvcmp}. Even for this ``worst case'' (where the artificial stars are 0.1 mag fainter than our magnitude limit), the color distribution of the artificial stars is much narrower than the observed one.
At all magnitudes and for all radial bins, more than 95\% of the synthetic stars were recovered.

\subsection{Radial trends in the HST/WFC3 data}
\label{sec:radial}

\begin{figure}
\includegraphics[width=\columnwidth]{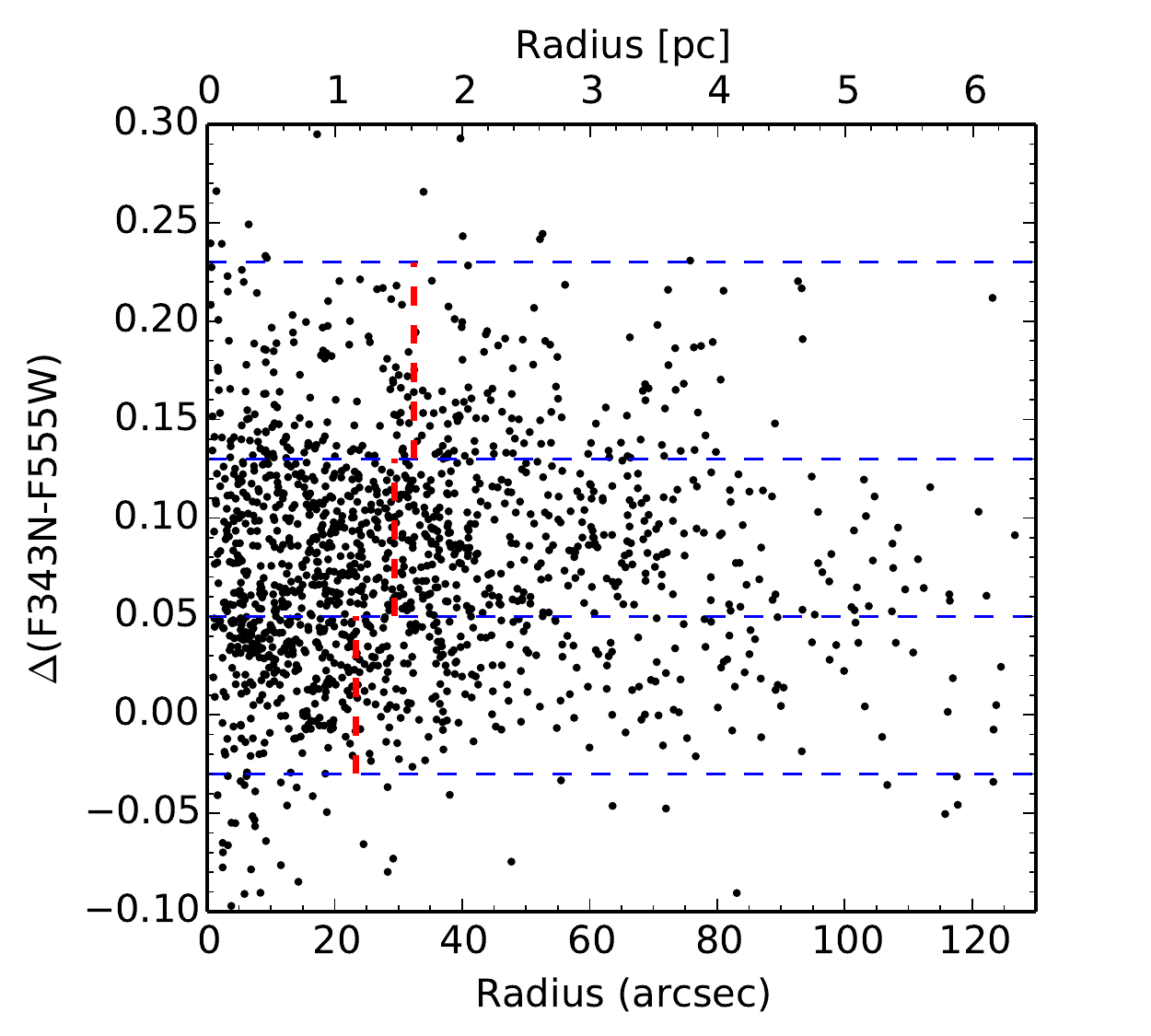}
\caption{\label{fig:m15rduv}The $\Delta$(F343N-F555W) color offset (relative to the ``Std composition'' isochrone in Fig.~\ref{fig:m15sel}) as a function of projected distance from the center of M15. The vertical (red) dashed lines indicate the median radial coordinate of each color group.}
\end{figure}

In Fig.~\ref{fig:m15rduv} we plot the observed $\Delta$(F343N-F555W) versus projected distance from the cluster center. The visual impression from this figure is that the stars with the \emph{bluest} $\Delta$(F343N-F555W) colors ($\Delta(\mathrm{F343N-F555W})\la0.05$) appear to be the \emph{most} centrally concentrated. There may also be a difference between stars with intermediate colors ($0.05 \la \Delta(\mathrm{F343N-F555W}) \la 0.13$) and the reddest (most N-enhanced) stars. 

Based on this first assessment of the data, we divided the  stars into ``N-normal'', ``intermediate'' and ``extreme'' samples, corresponding to
$-0.03<\Delta(\mathrm{F343N-F555W})\leq0.05$, 
$0.05<\Delta(\mathrm{F343N-F555W})\leq0.13$, and
$0.13<\Delta(\mathrm{F343N-F555W})<0.23$, respectively. To avoid contamination by blue stragglers, we further required $\mathrm{F555W}-\mathrm{F814W} > -0.6$ (see Figure~4 in Paper~I).
These groups contain 32\%, 50\%, and 18\% of the 1339 RGB stars in the magnitude range considered here and in the following we refer to them as groups A, B, and C. The different colors in Fig.~\ref{fig:m15sel} correspond to these three groups.
The dashed, vertical lines in Fig.~\ref{fig:m15rduv} show the median radial coordinates of each group; these are $R_\mathrm{med,A} = 23\farcs3\pm1\farcs1$, $R_\mathrm{med,B}=29\farcs4\pm1\farcs1$, and $R_\mathrm{med,C}=32\farcs4\pm2\farcs0$ for the three groups, respectively (errors were estimated via bootstrapping).
The group A stars are clearly more concentrated than the other groups. The artificial star tests showed that detection incompleteness is negligible over the mag\-ni\-tude- and radial ranges considered here. However, the \emph{spatial} completeness drops below 100\% at distances of $>40\arcsec$ from the cluster center and is only 10\% at our outer limit, 130\arcsec. While this should not affect the comparison of the different groups in a strictly relative sense, the absolute values of the median radii are therefore not related to physical cluster properties in a simple way.

Our A, B, and C groups are somewhat reminiscent of the primordial (P), intermediate (I) and extreme (E) populations defined by \citet{Carretta2009a} based on Na and O abundances. However, while Carretta et al.\ found a good correspondence between their spectroscopically defined populations and CN-sensitive photometry, it is not clear that their groups are exactly equivalent to ours. Indeed, they find no ``E'' stars in M15.
To avoid confusion, we therefore use a different naming scheme. We emphasize that the adopted division is not meant to imply the existence of three distinct populations. Indeed, there is no evidence for this in our CMD.  The photometry of \citet{Piotto2015} does suggest a bimodal structure of the RGB in M15, but also has a substantial number of stars with intermediate colors. \citet{Pancino2010a} found a bi-modal distribution of CN and CH band strengths from low-dispersion spectroscopy of main sequence stars in M15, while bimodality is less clear or absent in other data \citep{Kayser2008,Cohen2005a,Sneden1997}. 

\begin{figure}
\includegraphics[width=\columnwidth]{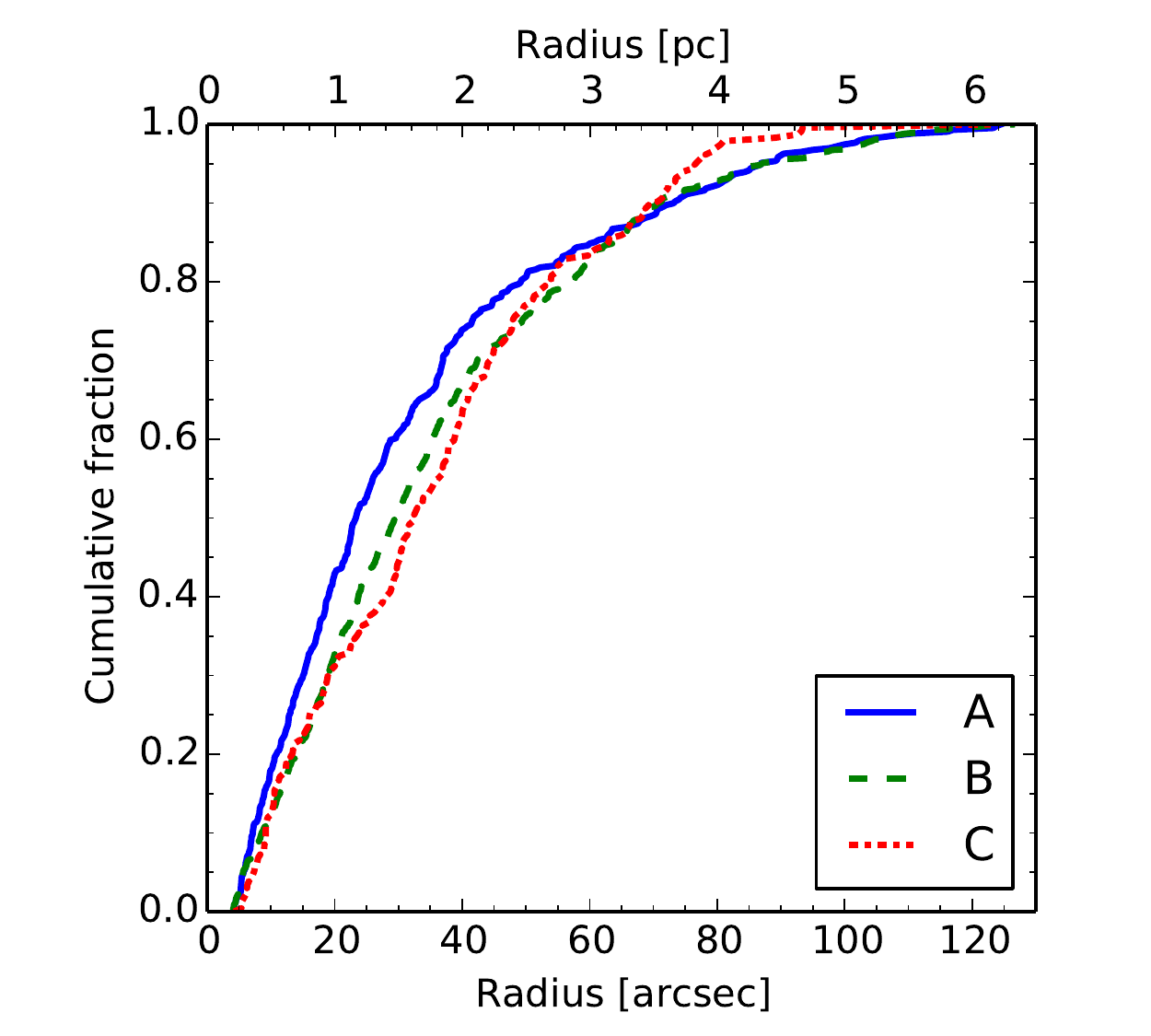}
\caption{\label{fig:m15rcum}Cumulative radial distributions of the populations. The primordial (group A) stars are clearly the most centrally concentrated.}
\end{figure}

Figure~\ref{fig:m15rcum} shows the cumulative radial distributions of the three groups. These confirm that the group A stars are the most centrally concentrated, followed by group B, and the group C stars are the least concentrated. From a K-S test, we find that the $P$ values when comparing the radial distributions of the group A and B stars, the A and C stars, and the B and C stars are $P_\mathrm{AB}=1.4\times10^{-3}$, $P_\mathrm{AC}=1.4\times10^{-5}$, and $P_\mathrm{BC}=0.19$, respectively. The A stars thus differ very significantly from both the group B and C stars, whereas the difference between the B and C stars is only marginally significant. This is consistent with the error estimates on the median radii above.

\begin{deluxetable*}{lccccccc}
\tablecaption{\label{tab:spatdat}Statistics of radial distributions for different selection criteria}
\tablecolumns{10}
\tablehead{
\colhead{Selection} & \colhead{$N$} & \colhead{$R_\mathrm{med,A}$} & \colhead{$R_\mathrm{med,B}$} & \colhead{$R_\mathrm{med,C}$} & \colhead{$P_\mathrm{AB}$} & \colhead{$P_\mathrm{AC}$} & \colhead{$P_\mathrm{BC}$}
}
\tabletypesize{\small}
\startdata
\multicolumn{3}{l}{Standard} \\
$+1 < \mathrm{F555W} < +3$ & 1339 & $23\farcs3\pm1\farcs2$ & $29\farcs4\pm1\farcs0$ & $32\farcs4\pm2\farcs0$ & $1.4\times10^{-3}$ & $1.4\times10^{-5}$ & 0.19 \\
$+1 < \mathrm{F555W} < +2.5$ & 761 & $22\farcs5\pm1\farcs4$ & $27\farcs8\pm1\farcs5$ & $31\farcs1\pm2\farcs2$ & 0.032 & $7.5\times10^{-4}$ & 0.23 \\
$+1 < \mathrm{F555W} < +2$ & 380 & $23\farcs0\pm2\farcs3$ & $28\farcs1\pm2\farcs0$ & $30\farcs5\pm2\farcs3$ & 0.40 & 0.11 & 0.68 \\
\multicolumn{3}{l}{With differential reddening correction} \\
$+1 < \mathrm{F555W} < +3$ & 1330 & $22\farcs7\pm1\farcs0$ & $29\farcs1\pm1\farcs0$ & $36\farcs8\pm2\farcs4$ & $2.4\times10^{-5}$ & $5.1\times10^{-9}$ & $1.0\times10^{-3}$ \\
$+1 < \mathrm{F555W} < +2.5$ & 760 & $22\farcs0\pm1\farcs6$ & $26\farcs3\pm1\farcs4$ & $33\farcs8\pm2\farcs2$ & $3.8\times10^{-3}$ & $3.1\times10^{-6}$ & $2.2\times10^{-3}$ \\
$+1 < \mathrm{F555W} < +2$ & 374 & $23\farcs1\pm2\farcs4$ & $26\farcs5\pm1\farcs9$ & $32\farcs3\pm2\farcs2$ & 0.071 & $5.2\times10^{-3}$ & 0.063 \\
\multicolumn{3}{l}{Removing blends} \\
$+1 < \mathrm{F555W} < +3$ & 1272 & $25\farcs1\pm1\farcs4$ & $30\farcs6\pm1\farcs0$ & $32\farcs8\pm2\farcs0$ & $6.4\times10^{-3}$ & $1.1\times10^{-4}$ & $0.27$ \\
$+1 < \mathrm{F555W} < +2.5$ & 730 & $23\farcs2\pm1\farcs6$ & $28\farcs3\pm1\farcs5$ & $31\farcs5\pm2\farcs4$ & $0.081$ & $1.9\times10^{-3}$ & $0.21$ \\
$+1 < \mathrm{F555W} < +2$ & 367 & $24\farcs8\pm2\farcs6$ & $28\farcs5\pm1\farcs9$ & $30\farcs6\pm2\farcs3$ & 0.49 & 0.19 & 0.71
\enddata
\tablecomments{``Standard analysis'' refers to the default procedure described in the main text, i.e.\ no differential reddening correction or explicit removal of blends.}
\end{deluxetable*}

These findings are robust to changes in the exact selection criteria and details of the analysis.
In Table~\ref{tab:spatdat} we list the median radii and $P$-values for different magnitude cuts and other modifications to our analysis procedure. 
If the correction for differential reddening is included, the differences become even more significant.
We also tried excluding blended stars, here defined as stars that have a neighbor within the fitting radius (3 pixels) that is brighter than $\mathrm{F555W}+2$, where the F555W magnitude refers to the magnitude of the star itself. This criterion removes about 5\% of the stars and the $P$ values increase somewhat. Because the blended stars are found preferentially in the inner regions of the cluster, the median radii all increase slightly, but the overall differences are preserved. 
For brighter magnitude limits, the number of stars decreases and with it the statistical significance of the differences, but it is always true that the group A stars are more centrally concentrated than the group B stars, which in turn are more concentrated than the group C stars.

\begin{deluxetable}{lcccc}
\tablecaption{\label{tab:spatdatr}Statistics of radial distributions split at ridge line}
\tablecolumns{5}
\tablehead{
\colhead{Selection} & \colhead{$N$} & \colhead{$R_\mathrm{med,-}$} & \colhead{$R_\mathrm{med,+}$} & \colhead{$P$}
}
\tabletypesize{\small}
\startdata
\multicolumn{3}{l}{Standard} \\
$+1 < \mathrm{F555W} < +3$ & 1364 & $25\farcs0 \pm 1\farcs1$ & $30\farcs6 \pm 0\farcs9$ & $3.6\times10^{-4}$ \\
$+1 < \mathrm{F555W} < +2.5$ & 769 & $23\farcs7 \pm 1\farcs2$ & $29\farcs1 \pm 1\farcs2$ & 0.017 \\
$+1 < \mathrm{F555W} < +2$ & 381 & $25\farcs3 \pm 2\farcs3$ & $28\farcs6 \pm 1\farcs4$ & 0.50 \\
\multicolumn{3}{l}{With differential reddening correction} \\
$+1 < \mathrm{F555W} < +3$ & 1368 & $23\farcs4 \pm 0\farcs9$ & $30\farcs9 \pm 0\farcs8$ & $2.7\times10^{-7}$ \\
$+1 < \mathrm{F555W} < +2.5$ & 784 & $22\farcs6 \pm 1\farcs1$ & $30\farcs0 \pm 1\farcs2$ & $1.5\times10^{-4}$ \\
$+1 < \mathrm{F555W} < +2$ & 386 & $23\farcs8 \pm 1\farcs7$ & $30\farcs1 \pm 1\farcs2$ & 0.021 \\
\multicolumn{3}{l}{Removing blends}  \\
$+1 < \mathrm{F555W} < +3$ & 1292 & $26\farcs4 \pm 1\farcs3$ & $31\farcs6 \pm 1\farcs1$ & $9.1\times10^{-4}$ \\
$+1 < \mathrm{F555W} < +2.5$ & 738 & $24\farcs9 \pm 1\farcs3$ & $29\farcs8 \pm 1\farcs2$ & $0.024$ \\
$+1 < \mathrm{F555W} < +2$ & 368 & $26\farcs0 \pm 2\farcs6$ & $28\farcs8 \pm 1\farcs3$ & 0.46
\enddata
\tablecomments{The columns $R_\mathrm{med,-}$ and $R_\mathrm{med,+}$ give the median projected distances from the center of M15 for stars to the left and right of the empirical ridge line, respectively.}
\end{deluxetable}

Table~\ref{tab:spatdatr} shows the statistics of the radial distributions when the RGB stars are divided according to an empirical ridge line in the F343N-F555W vs.\ F555W diagram, instead of using the theoretical models as a reference. The ridge line (shown as a thick black line in Fig.~\ref{fig:m15sel}) was defined similarly to the ridge line for main sequence stars used for the reddening map, and the RGB stars were divided into two groups with bluer and redder colors than the ridge line, respectively.
As in Table~\ref{tab:spatdat}, the difference between the radial distributions of the two sub-samples is highly significant.  
For the ``standard analysis'', 
the median radii for the blue and red populations are $R_\mathrm{med,-} = 25\farcs0\pm1\farcs1$ and $R_\mathrm{med,+} = 30\farcs6\pm0\farcs9$ with $P=3.6\times10^{-4}$ (here, we use `$-$' and `$+$' to denote the stars that are bluer and redder than the ridge line).
For the brighter bins, the difference is again less significant because of the smaller number of stars, but numerically it remains consistent with the bins that include fainter stars and thus have better statistics.

\begin{figure}
\includegraphics[width=\columnwidth]{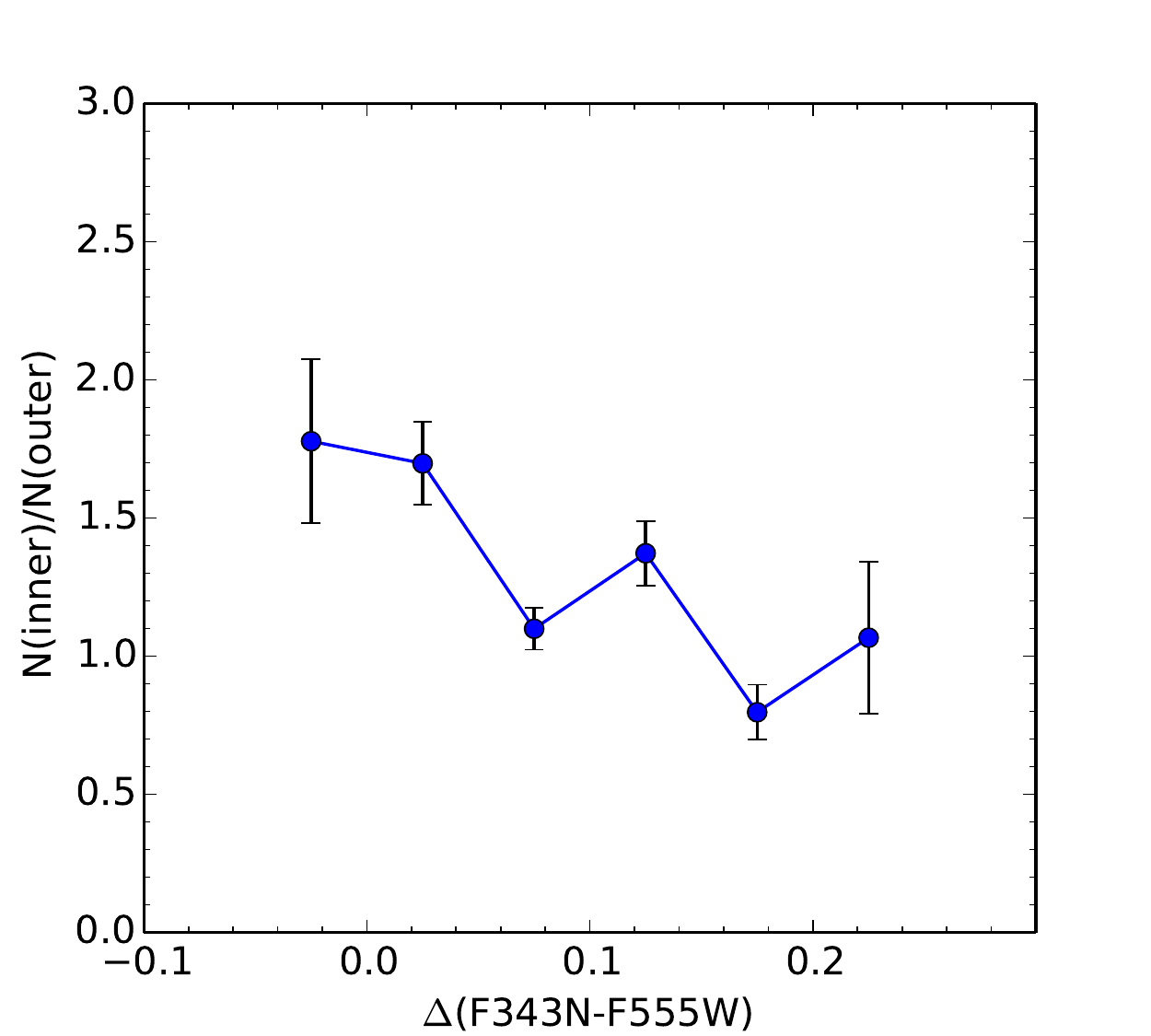}
\caption{\label{fig:m15rio}Ratio of color distributions in the inner ($R<20\arcsec$) and outer ($R>20\arcsec$) regions of M15. The inner regions preferentially contain stars with blue F343N-F555W colors.}
\end{figure}

An alternative way of illustrating the differences in the radial distributions is shown in Fig.~\ref{fig:m15rio}, where we plot the number ratio of stars in the inner regions ($R<20\arcsec$) vs.\ the outer regions ($R>20\arcsec$) as a function of $\Delta(\mathrm{F343N-F555W})$. We see that stars with blue $\Delta(\mathrm{F343N-F555W})$ colors are more prevalent near the center, which is consistent with the differences in the cumulative radial distributions. The difference between the color distributions in the inner and outer regions is again highly significant, with $P=6.6\times10^{-4}$.

\begin{figure}
\includegraphics[width=\columnwidth]{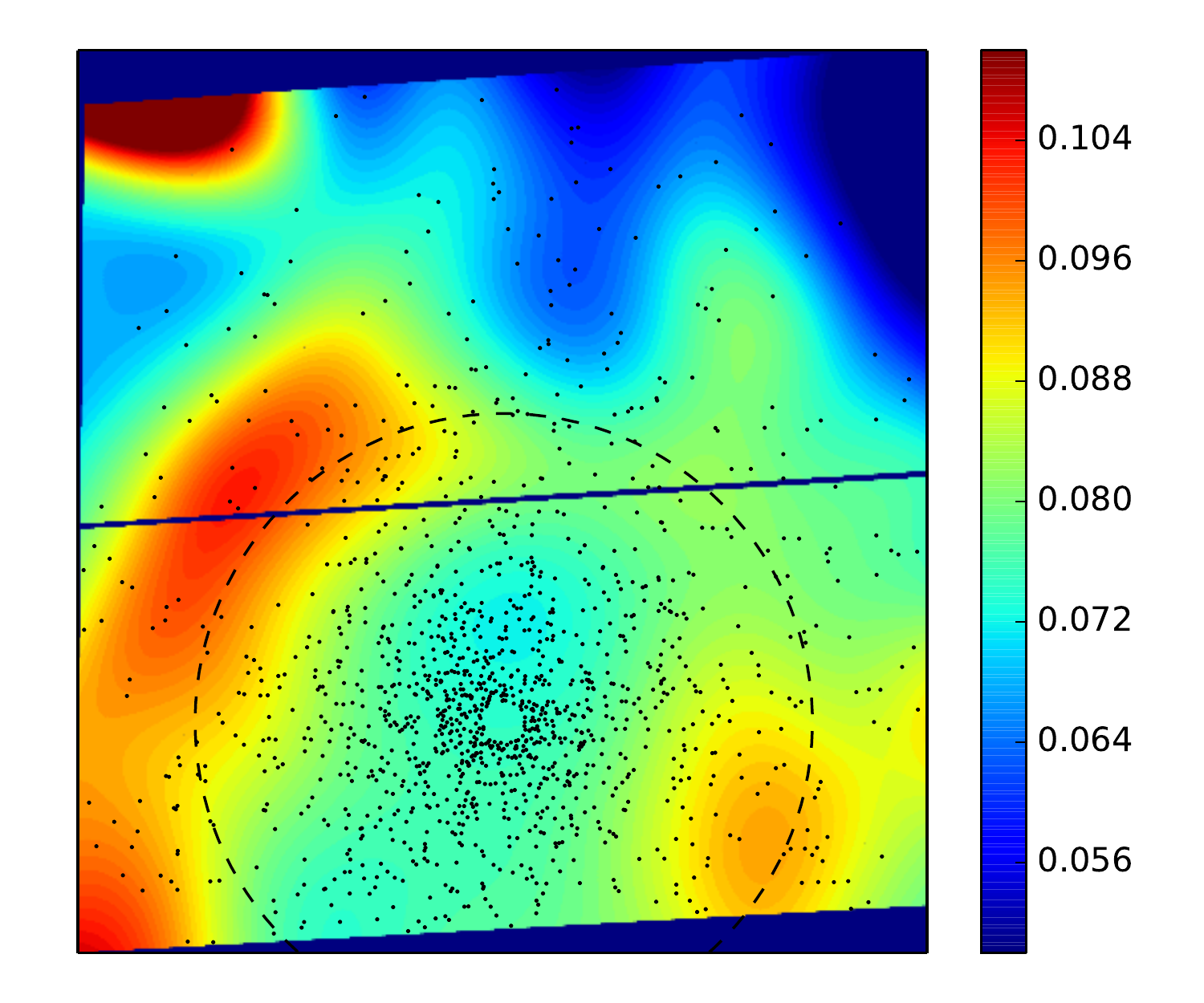}
\caption{\label{fig:duvmap}Map of the $\Delta(\mathrm{F343N}-\mathrm{F555W})$ color of RGB stars (shown with black points). The mean colors are bluer near the center of M15, in agreement with the inferred higher fraction of N-normal stars there. The dashed circle is centered on M15 and has a radius of $1\arcmin$. }
\end{figure}

As discussed in Sect.~\ref{sec:dred}, there may be small variations in the reddening and photometric zero-points across the field. Could such variations cause the observed trends? We have already noted that a correction for differential reddening actually \emph{increases} the statistical significance of the trends. However, it is uncertain to what extent the structure in the reddening map is real. Figure~\ref{fig:duvmap} shows a map of the average $\Delta$(F343N-F555W) color across the WFC3 field. The map was produced in the same way as the reddening map in Fig.~\ref{fig:extmap}, except that a larger smoothing radius (350 pixels) was used because of the smaller number of RGB stars. 
The mean $\Delta$(F343N-F555W) colors are clearly bluer near the center of M15, which is consistent with the radial trends in Figs.~\ref{fig:m15rcum} and \ref{fig:m15rio}. At radial distances of $\sim1\arcmin$ from the center, the $\Delta$(F343N-F555W) colors are 0.02-0.03 mag redder than at the center. The reddest colors are seen in the lower left-hand corner, where Fig.~\ref{fig:extmap} also shows the highest reddening, which may suggest that there is a real gradient in the reddening across the field. However, the general morphology of Fig.~\ref{fig:duvmap} is quite different from that of Fig.~\ref{fig:extmap}, and the former appears much more symmetric with respect to the center of M15, especially if one imagines subtracting an overall left-right gradient. If the color variations were dominated by differential reddening, we would expect a greater degree of similarity between the two maps, so it appears likely that the F343N-F555W color variations are, for the most part, intrinsic to M15. However, with our present data it remains difficult to exclude with certainty the possibility that some combination of small reddening variations and zero-point variations could affect the observed radial distributions noticeably. 

\subsection{Combining the HST data with SDSS photometry}

Because of the limited field of view, the WFC3 data only allow us to constrain radial trends out to about 130\arcsec, with poor statistics in the outer parts because of incomplete spatial coverage. However, \citet{An2008} have carried out PSF-fitting photometry on images from the \textit{Sloan Digital Sky Survey} \citep[SDSS;][]{York2000} for a number of fields around Galactic globular clusters, including M15.  While these data lack the spatial resolution of HST imaging and cannot resolve the inner parts of the clusters, they  extend to much larger radii. The Sloan $u$ filter covers NH and CN absorption bands, which makes the $u-g$ color sensitive to light element abundances.
The \citet{An2008} photometry was analyzed by \citet{Lardo2010}, who found that stars with redder $u-g$ colors (i.e., the enriched stars) tended to be more radially concentrated in M15 (and other GCs). This trend is \emph{opposite} to that seen in Fig.~\ref{fig:m15rcum} from the WFC3 data in the central regions of M15. These results are, however, not necessarily in contradiction to each other, because of the limited overlap of the radial ranges covered by the two datasets.

\begin{figure}
\includegraphics[width=\columnwidth]{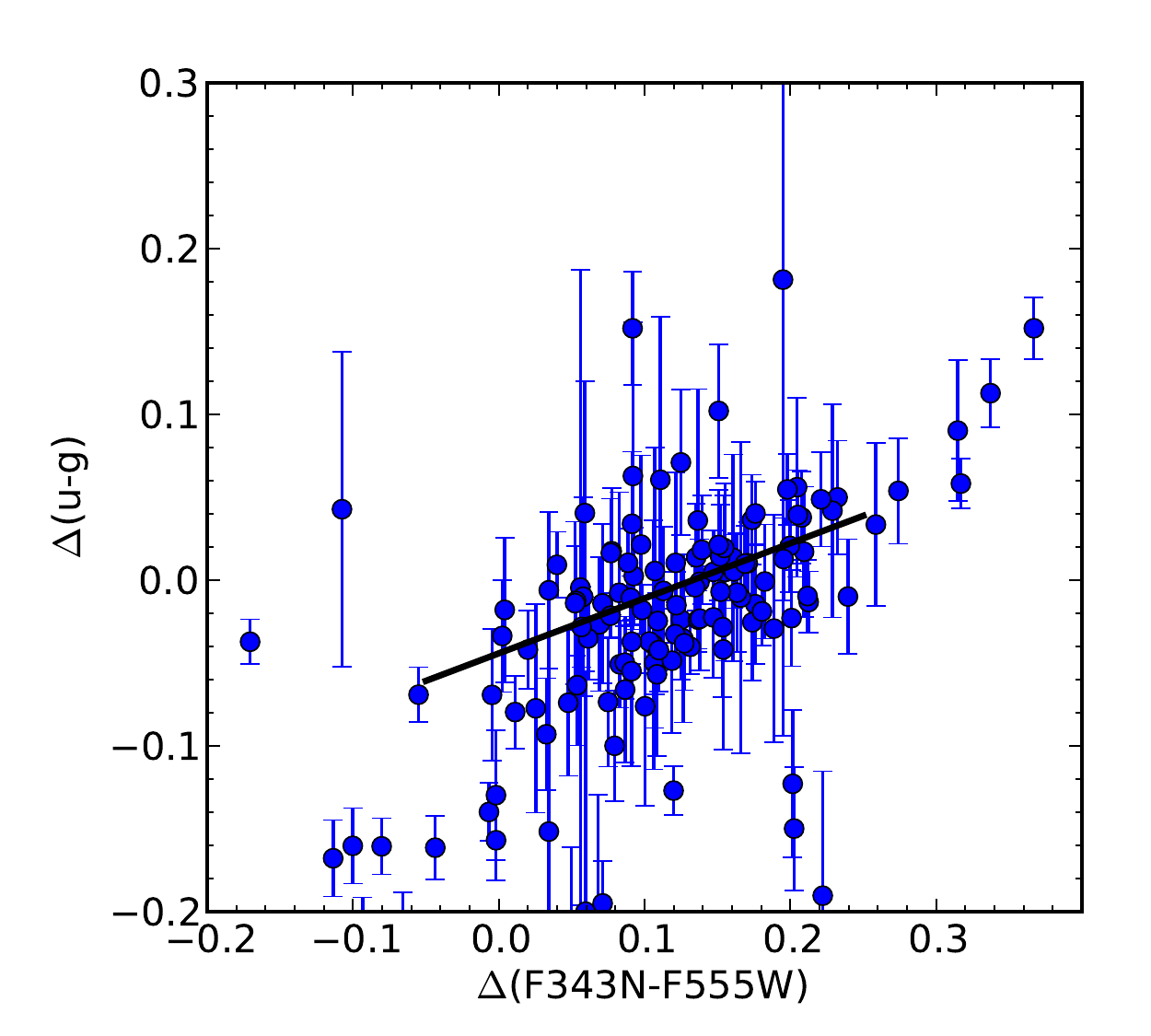}
\caption{\label{fig:hst-sdss}Comparison of the SDSS $\Delta(u-g)$ and HST $\Delta$(F343N-F555W) color offsets.}
\end{figure}

To explore the variation in the ratio of primordial vs.\ enriched stars over the full radial range, we combined our HST data with the SDSS photometry of \citet{An2008}. The SDSS data extend to about $20\arcmin$ from the center of M15, but following \citet{Lardo2010} we restrict our analysis to the innermost $10\arcmin$ to limit contamination by field stars.
The filter combinations of the two datasets are similar, but not identical, so we first tested whether they yield consistent results in the region of overlap (at radii of 60\arcsec--130\arcsec). We defined a $\Delta(u-g)$ parameter for the SDSS data in the same way as the $\Delta$(F343N-F555W) parameter. As a reference, we used the same isochrone employed in Fig.~\ref{fig:m15sel} with the SDSS colors provided through the Dartmouth web interface\footnote{\texttt{stellar.dartmouth.edu/\~{}models/isolf.html}}. Our $\Delta(u-g)$ parameter is closely analogous to the $\Delta_{u-g}$ parameter defined by Lardo et al., the main difference being that they used an empirical ridge line as a reference whereas we use a theoretical isochrone. In Fig.~\ref{fig:hst-sdss} we compare the $\Delta$(F343N-F555W) and $\Delta(u-g)$ measurements for stars in common between the HST and SDSS datasets. 
For the SDSS data we adopted the $15 < g < 17$ magnitude range of \citet{Lardo2010}, where the faint limit is equivalent to F555W $\approx16.6$ or $M_\mathrm{F555W}\approx1.24$. 
Since this is close to the bright magnitude cut-off in our HST data, we relaxed the bright magnitude limit in the HST data to $M_\mathrm{F555W} = +0.5$ (for the purpose of this comparison only) to increase the overlap between the two samples, and we also eliminated the $\chi^2_\nu$ cut. As noted by \citet{Lardo2010}, the photometric errors account for a significant fraction of the dispersion in $u-g$ and there is indeed a substantial scatter in Fig.~\ref{fig:hst-sdss}. 
Nevertheless, there is a significant correlation between $\Delta$(F343N-F555W) and $\Delta(u-g)$, with the straight line showing a linear least-squares fit to stars with $-0.1 < \Delta(u-g) < 0.1$ and $-0.05 < \Delta(\mathrm{F343N}-\mathrm{F555W}) < 0.25$. The fit yields $\Delta(u-g) = (0.33\pm0.07) \times \Delta(\mathrm{F343N}-\mathrm{F555W}) - (0.042\pm0.009)$.

For comparison with the fit in Fig.~\ref{fig:hst-sdss}, we used \texttt{ATLAS12}/\texttt{SYNTHE} synthetic spectra to calculate the expected color difference between N-normal and N-enhanced models for the SDSS filters, following the same approach as in Paper~I and for the models in Fig.~\ref{fig:colcmp}. For stars on the lower RGB, we found a difference of 0.05 mag in $\Delta(u-g)$, i.e.\ 0.31 times the 0.16 mag difference in $\Delta$(F343N-F555W). There is thus a very good agreement between the theoretical (0.31) and measured ($0.33\pm0.07$) slopes of the $\Delta(u-g)$ vs.\ $\Delta$(F343N-F555W) relation. We therefore used the linear fit in Fig.~\ref{fig:hst-sdss} to transform the $\Delta(u-g)$ values to $\Delta$(F343N-F555W) values.
To make a clear distinction between $\Delta$(F343N-F555W) values transformed from the SDSS photometry and those measured directly in the HST data, we will denote the former by $\Delta$(F343N-F555W)$_T$ in the following.

\begin{figure}
\includegraphics[width=\columnwidth]{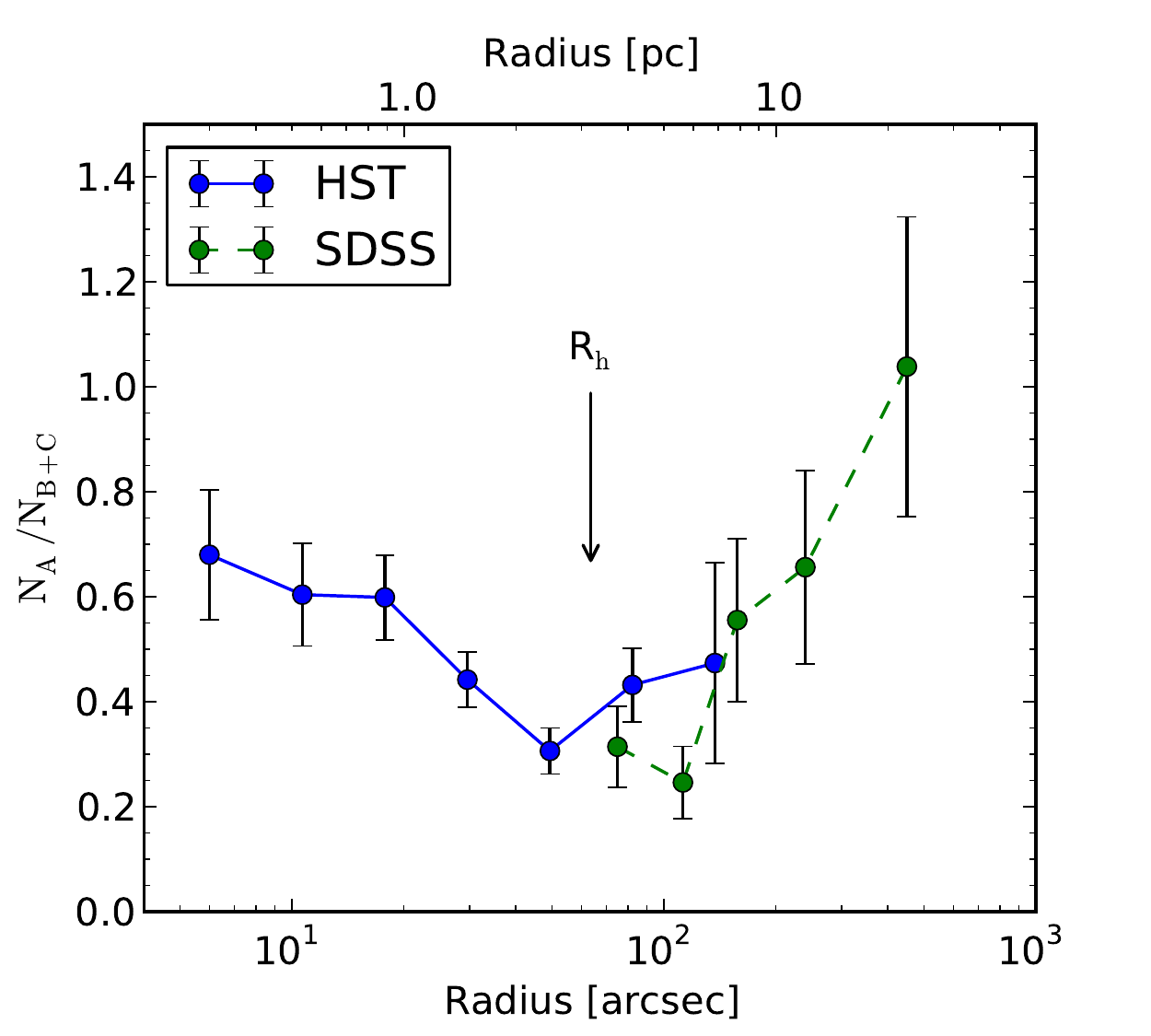}
\caption{\label{fig:m15rmrg}Ratio of primordial vs.\ enriched stars as a function of radial distance. HST and SDSS datasets have been combined to cover the radial range from $4\arcsec$ to $10\arcmin$. The arrow indicates the half-light radius, $R_h=1\farcm06$ \citep{Harris1996}.}
\end{figure}

Figure~\ref{fig:m15rmrg} shows the number ratio $N_A/N_{B+C}$ as a function of radius for the combined HST/WFC3 and SDSS data, where $N_A$ is the number of group A stars. We count the group B and C stars together, $N_{B+C}$, since these cannot be well distinguished in the SDSS photometry, and the difference between the spatial distributions of these two groups is the least significant according to the HST data. To account for the larger scatter in the SDSS data, we have extended the red and blue limits of the $\Delta$(F343N-F555W)$_T$ range by 0.05 mag at both the red and blue edge, thus including stars in the range $-0.08 < \Delta(\mathrm{F343N-F555W})_T<0.28$. We have applied the same color cut at $\Delta(\mathrm{F343N-F555W})_T=0.05$ between group A and B$+$C as in the preceding figures. Comparing with the photometry at larger radii, we found that contamination in the SDSS sample is limited to $<10$\% after applying these selection criteria, even in the outermost bin.

From Fig.~\ref{fig:m15rmrg}, the WFC3 and SDSS samples agree reasonably well on the $N_A/N_{B+C}$ ratio in the overlap region. At radii $\la 60\arcsec$, the $N_A/N_{B+C}$ ratio increases towards the center, consistent with the more centrally concentrated distribution of the group A stars seen in Figs.~\ref{fig:m15rduv} and \ref{fig:m15rcum}. Around one arcminute (3 pc) from the center, the $N_A/N_{B+C}$ ratio has a \emph{minimum} and then increases again towards larger radii, in agreement with the analysis by \citet{Lardo2010}. A hint of this increase is seen already in the WFC3 photometry (although the outermost bin has a large error bar) and from the ``donot-shaped'' morphology of Fig.~\ref{fig:duvmap}, but it becomes very clear in the combined dataset.
\emph{We thus find that the $N_A/N_{B+C}$ ratio has a U-shaped dependency on radius with a minimum near the half-light radius} \citep[$1\farcm06$;][]{Harris1996}. This is our key result.

We note that the brightest stars in the SDSS photometry are above the threshold where \citet{Gratton2000} find evidence of deep mixing. These authors find modified abundance patterns for $\log L/L_\odot \ga 1.8$ in metal-poor field giants, corresponding to $M_g \la 0.8$, whereas the absolute magnitude range of the M15 SDSS photometry is $-0.42 < M_g < 1.58$. Thus from Figure~10 of \citet{Gratton2000}, deep mixing may have enhanced the N abundances by up to $\sim0.5$ dex in the brightest giants in the SDSS data. This is still a relatively small effect compared to the overall $\sim2$ dex variations between the different populations and, moreover, the effect would be to move stars from the group A to the group B$+$C bins and thereby lead us to \emph{underestimate} the true $N_A/N_{B+C}$ ratio from the SDSS photometry.

\subsection{The effect of mass segregation on populations with different He abundance: $N$-body simulations}
\label{sec:nbody}

A tendency for the enriched stars to prefer a location near the half-mass radius is unexpected in all current scenarios for the origin of multiple stellar populations in globular clusters. 
A possible explanation could be a significantly enhanced He abundance of the enriched stars and modification of the initial radial distributions by two-body relaxation. Because of the He enhancement, the enriched stars 
would undergo faster stellar evolution so that the enriched (group B$+$C) post-main sequence stars have lower masses than the primordial (group A) ones for a given age. Mass segregation due to two-body relaxation would then push the enriched giants outwards, decreasing their number fraction
in the centre. In the outer cluster parts the relaxation time is much longer, preserving the initial ratio for a much longer time, 
hence one might expect to find a U-shaped profile in the number ratio of the populations similar to that seen in Fig.~\ref{fig:m15rmrg}.

\begin{figure}
\includegraphics[width=\columnwidth]{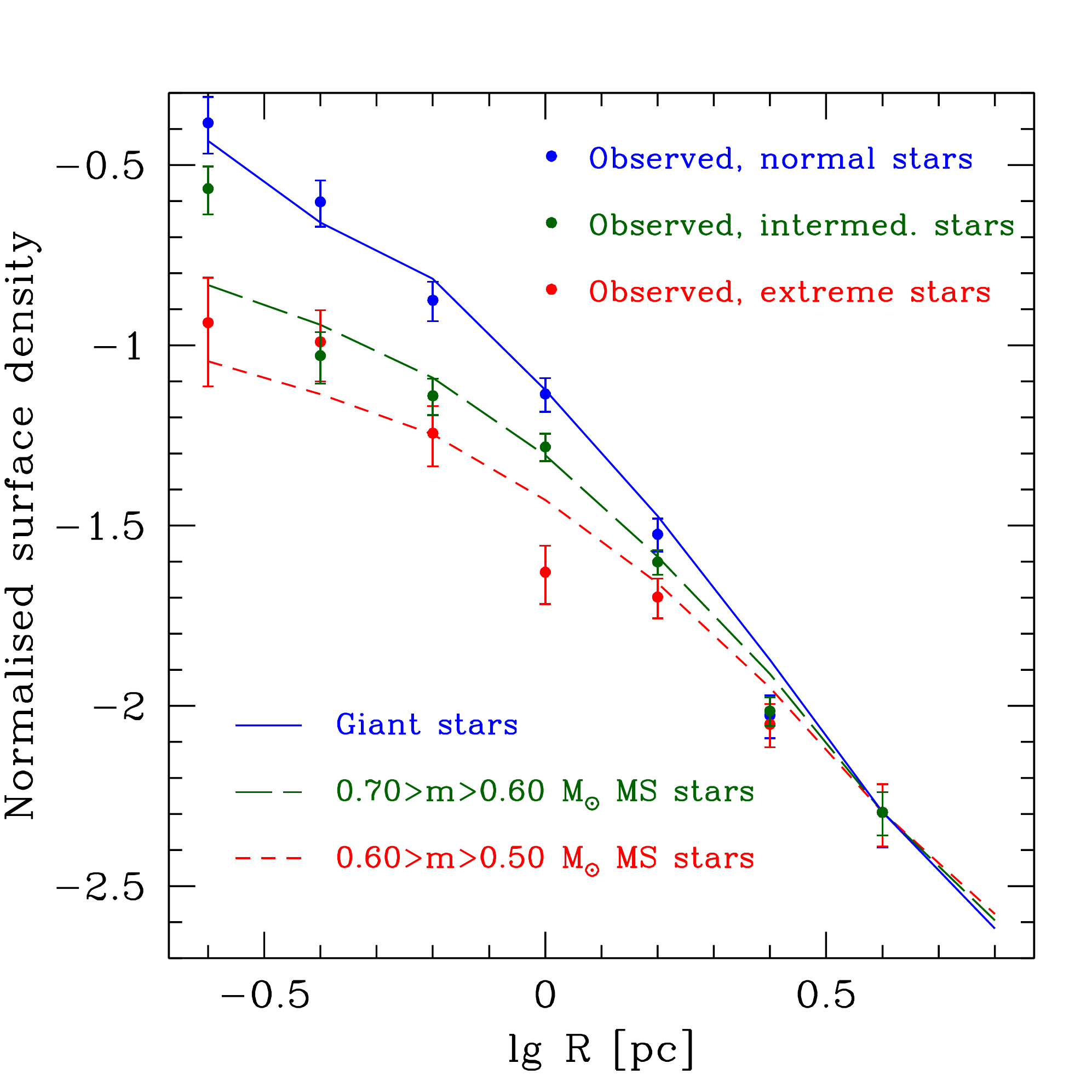}
\caption{\label{fig:m15seg}Observed radial distributions of N-normal stars and N-enhanced stars, compared with $N$-body simulations for stars of different masses that start with the same radial distribution. The curve labeled `Giant stars' is for masses of $0.82 \, M_\odot$. The lines show the simulated radial distributions after 11.5 Gyr, normalized to the outermost measured point.}
\end{figure}

In order to test what mass differences between primordial and enriched stars are necessary to explain the
radial trends seen in M15, we used the grid of $N$-body simulations used by \citet{McNamara2012} and \citet{Lutzgendorf2013} and selected the best-fitting non-IMBH cluster from this grid. Fig.~\ref{fig:m15seg} shows the radial distribution of stars with different masses at T=11.5 Gyr after all clusters were scaled to the mass and size of M15. The ``giant stars'' have masses of $0.82 \, M_\odot$ and the lower mass stars have been arranged into bins of 0.50 $M_\odot$ -- 0.60 $M_\odot$ and 0.60 $M_\odot$ -- 0.70 $M_\odot$. Because the $N$-body simulations of \citet{McNamara2012} start without mass segregation, we cannot expect to fit the outer profile of N-enhanced stars in M15, therefore we restrict our fit to radii inside the half-mass radius. Because of the long relaxation time in the outer regions, mass segregation due to two-body relaxation will not have developed there within a Hubble time, and a dedicated set of $N$-body simulations including initial segregation of the different populations would be required to reproduce the full radial profile.

As can be seen, the radial distribution of the giant stars in the $N$-body simulation provides a very 
good fit to the distribution of the N-normal giants in M15 (here, the observed radial distributions have been corrected for spatial incompleteness). The radial distribution of the most N-enhanced (group C) stars is best fitted
by the distribution of stars with masses between $0.50 \, M_\odot < M < 0.60 \, M_\odot$ in the $N$-body simulation, whereas the intermediate stars (group B) correspond better to masses of $0.60 \, M_\odot < M < 0.70 \, M_\odot$. Hence 
if the different radial distributions are due to two-body relaxation driven mass segregation, then the most enriched RGB stars would need to have masses of about $0.25 \, M_\odot$ less than the primordial giants. This requires an extreme degree of helium enrichment: for example, according to the models of \citet{Dotter2007},  $Y = 0.40$ instead of $Y = 0.25$ produces a mass difference of 0.20 $M_\odot$ for stars on the lower RGB at an age of 13 Gyr ($M = 0.79 \, M_\odot$ for $Y=0.25$ and $M = 0.59 \, M_\odot$ for $Y=0.40$). Because the initial conditions of the simulations are not mass segregated, the required mass differences (and, consequently, the required He enhancement) may be considered conservative estimates. If the cluster started out with a more concentrated enriched population, an even larger difference in mass between the giants of the two populations would be required to reverse this and produce the observed differences in the available time.

\subsection{Observational constraints on He abundance variations in M15}

A significant enhancement of the He abundance in the enriched stars is expected in most scenarios, due to the H-burning nucleosynthesis involved
\citep{Ventura2001,DAntona2002,Decressin2007,DeMink2009}. The observations of distinct, parallel main sequences that cannot be explained by differences in heavy element abundances provide strong evidence for He abundance variations in some GCs. In some clusters, such as $\omega$ Cen and NGC~2808, the He fraction of the enriched stars may be as high as $Y\sim0.40$ \citep{Bedin2004,Norris2004,Piotto2007}, but in others 
the enhancement is much more modest; photometry of the clusters 47 Tuc, NGC~6397, and NGC~6752 indicates He enhancement in the range $\Delta Y = 0.01$--0.03 \citep{Milone2012c,Milone2012a,Milone2013}. 
It is difficult to measure He abundances directly from spectroscopy of cool stars because of the large difference between the two lowest energy levels of neutral He. In hot horizontal branch stars, where He is more readily measurable, the surface composition may have been heavily modified by stellar evolutionary effects \citep{Behr2000,Valcarce2014}. Star-to-star differences in the strength of the chromospheric He {\sc i} 10\,830 \AA\ line in red giants have been observed in $\omega$ Cen and NGC~2808 and imply He abundances consistent with the large variations derived from photometry \citep{Dupree2011,Pasquini2011}, but deriving accurate abundances from the line is difficult. 
Unfortunately, neither high-precision photometry of sufficient depth to reveal multiple main sequences nor spectroscopic constraints on the He abundance of red giants currently exist for M15.

\begin{figure}
\includegraphics[width=80mm]{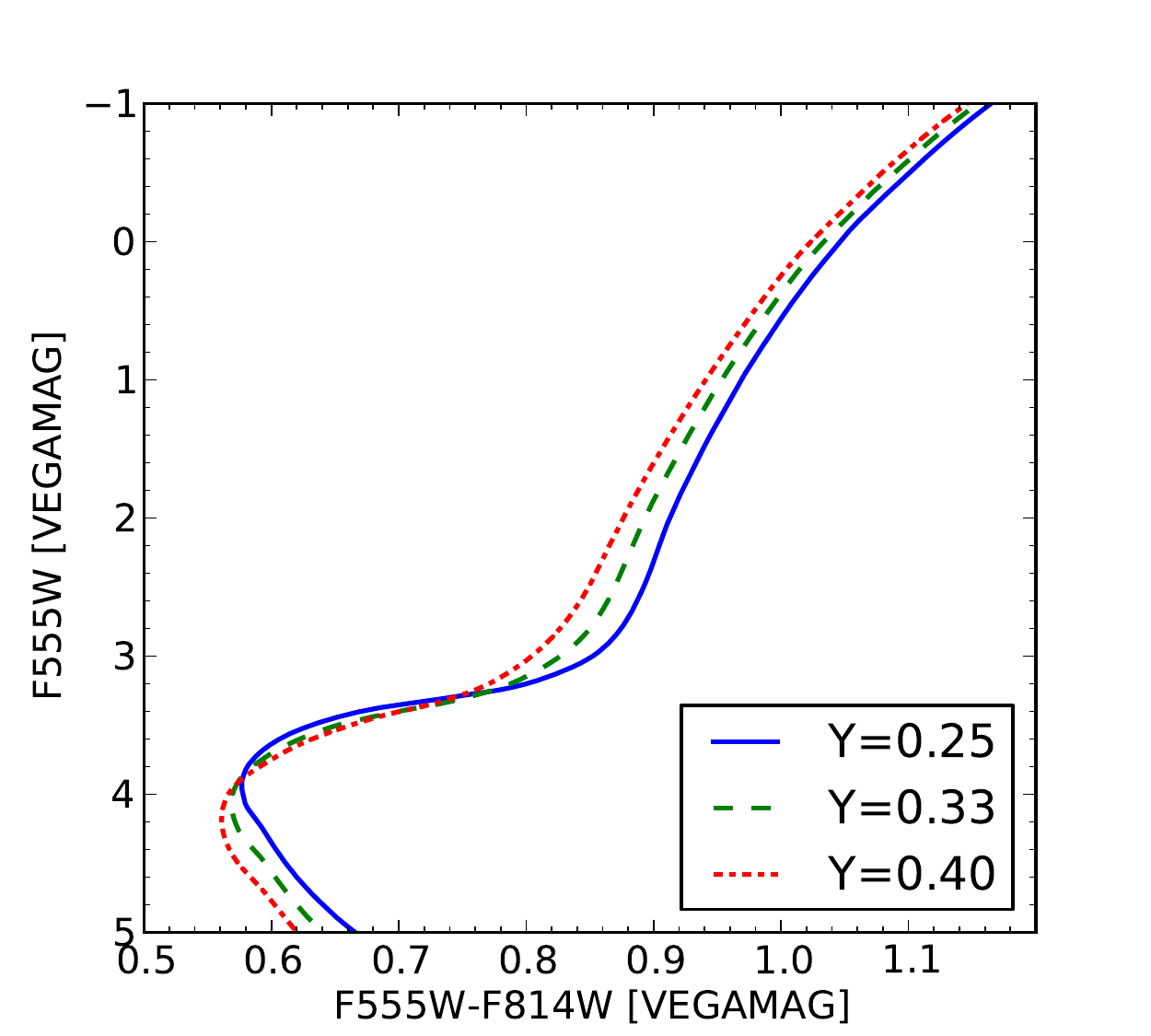}
\includegraphics[width=80mm]{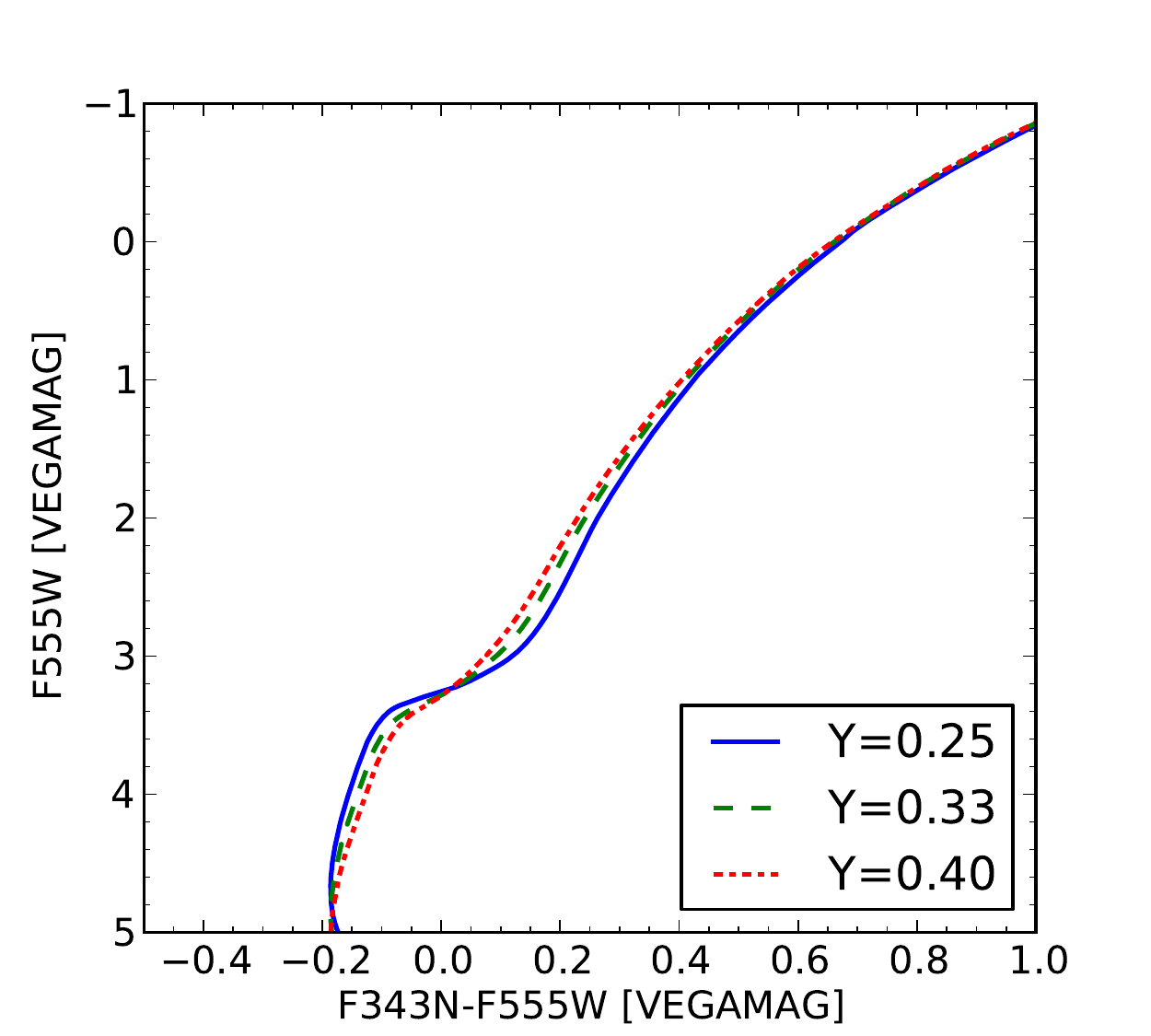}
\caption{\label{fig:hecmp}\citet{Dotter2007} isochrones for different He abundances. The isochrones have an age of 13 Gyr and the same $Z$, corresponding to $\mathrm{[Fe/H]}=-2.3$, $\mathrm{[Fe/H]}=-2.25$, and $\mathrm{[Fe/H]}=-2.20$ for $Y=0.25$, $Y=0.33$, and $Y=0.40$, respectively.}
\end{figure}

As noted in Sec.~\ref{sec:nvar}, not only the main sequence, but also the effective temperatures of stars on the lower part of the RGB should be sensitive to He abundance \citep{Salaris2006,DiCriscienzo2011,Beccari2013}.  Fig.~\ref{fig:hecmp} shows \citet{Dotter2007} isochrones for an age of 13 Gyr and $[\alpha/\mathrm{Fe}]=0.4$ and
 different He abundances ($Y=0.25$, 0.33, and 0.40). Here we have scaled $\mathrm{[Fe/H]}$ for $Y=0.33$ and $Y=0.40$ to the same total $Z$ (i.e., $\mathrm{[Fe/H]}=-2.25$ for $Y=0.33$ and $\mathrm{[Fe/H]}=-2.2$ for $Y=0.40$). 
In both F555W-F814W and F343N-F555W, the colors of RGB stars are predicted to be bluer for the He-enhanced models (keeping the light element mixture fixed). At $M_\mathrm{F555W}=+2$, the difference between the $Y=0.25$ and $Y=0.40$ models is $-0.036$ mag in F555W-F814W and $-0.034$ mag in F343N-F555W (for $Y=0.33$ vs.\ $Y=0.25$, the corresponding differences are $-0.020$ mag in F555W-F814W and $-0.021$ mag in F343N-F555W). If we instead keep the iron abundance relative to hydrogen fixed at $\mathrm{[Fe/H]}=-2.3$, then the F555W-F814W color offsets remain the same as above, whereas the difference between the $Y=0.25$ and $Y=0.40$ colors now amounts to $-0.050$ mag in $\mathrm{F343N}-\mathrm{F555W}$. In any case, the observed spread in F343N-F555W cannot be explained by He abundance variations and remains dominated by N abundance variations (cf.\ Fig.~\ref{fig:colcmp}). For F555W-F814W, however, the situation is the opposite: this color is largely insensitive to N abundance and any difference in F555W-F814W would therefore be attributable to He abundance variations. 

\begin{figure}
\includegraphics[width=\columnwidth]{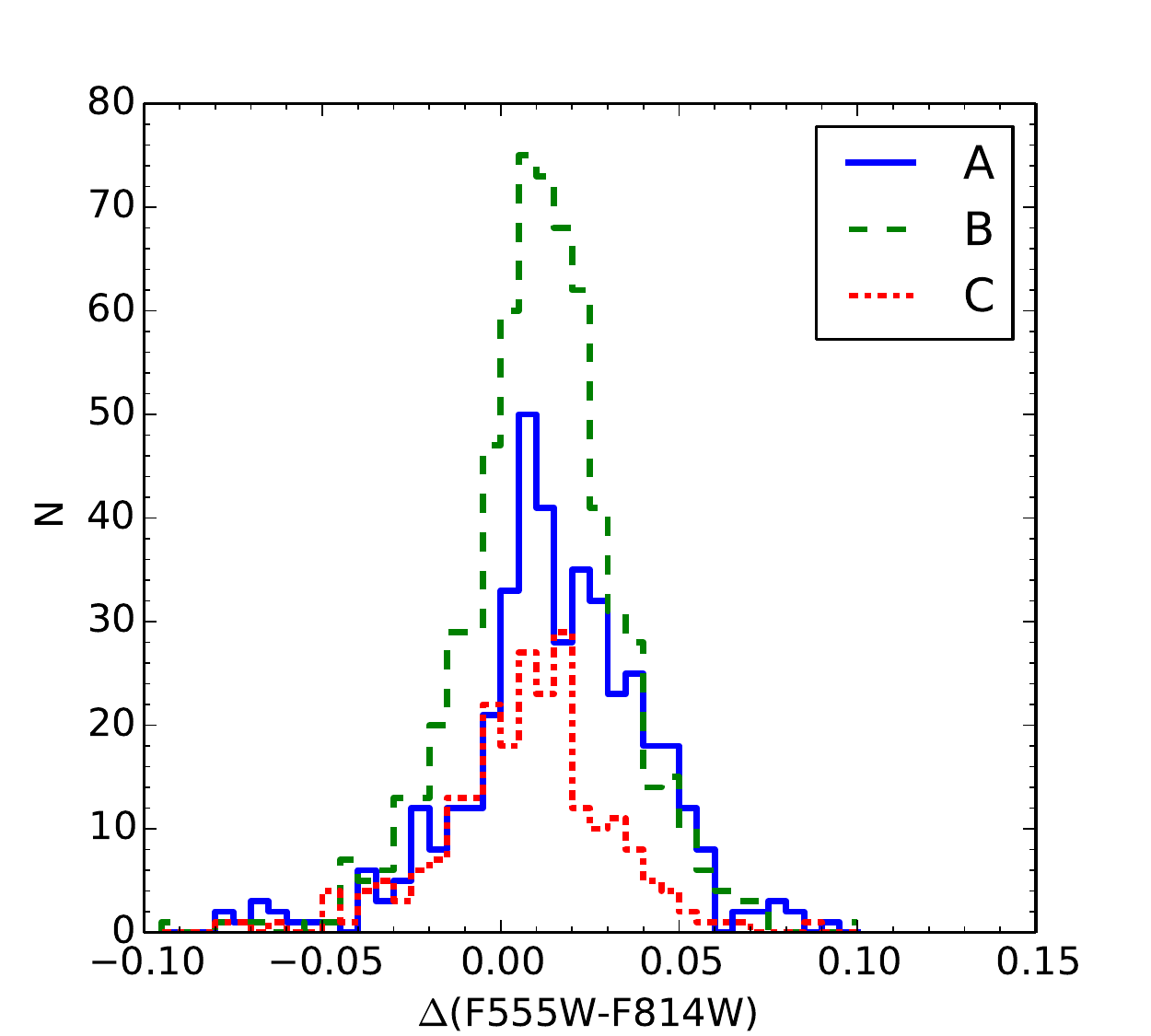}
\caption{\label{fig:m15vihist}Distributions of $\Delta$(F555W-F814W) for group A, B, and C stars.}
\end{figure}

In Fig.~\ref{fig:m15vihist} we show the observed $\Delta$(F555W-F814W) distributions for the group A, B, and C stars. 
The mean colors of the three groups are
$\langle \Delta(\mathrm{F555W}-\mathrm{F814W})\rangle_A = 0.015$ mag, 
$\langle \Delta(\mathrm{F555W}-\mathrm{F814W})\rangle_B = 0.011$ mag, and
$\langle \Delta(\mathrm{F555W}-\mathrm{F814W})\rangle_C = 0.007$ mag. 
There is, indeed, a tendency for the group B and C stars to have bluer F555W-F814W colors than the group A stars, as would be expected if there is a difference in He abundance.
From Fig.~\ref{fig:hecmp}, we get $\Delta$(F555W-F814W)/$\Delta Y = 0.24$ (at $M_\mathrm{F555W}=2$), so a color difference of 0.008 mag between the group A and C stars corresponds to $\Delta Y \sim 0.033$. The corresponding mass difference on the RGB is only $0.045 \, M_\odot$, which is much too small to explain the differences in spatial distribution as a consequence of mass segregation. The differences in $\langle \Delta(\mathrm{F555W}-\mathrm{F814W})\rangle$ between the groups become even smaller ($<0.002$ mag) if the differential reddening correction is applied.

\begin{figure}
\includegraphics[width=\columnwidth]{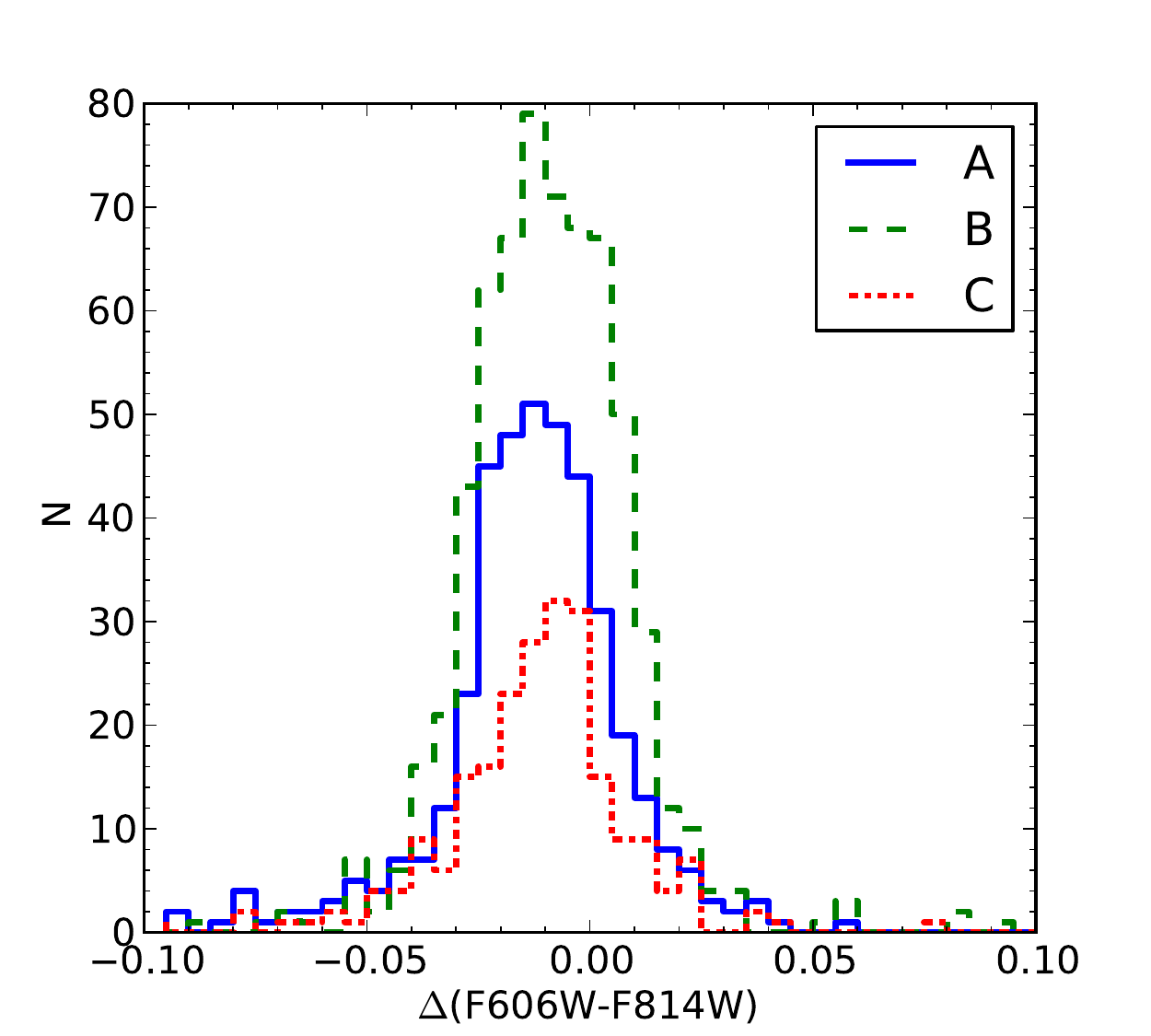}
\caption{\label{fig:acshist}Distributions of $\Delta$(F606W-F814W) for group A, B, and C stars. There are no significant differences between the three groups.}
\end{figure}

As an independent check, we made the same comparison using the ACSGCS F606W-F814W colors. Unlike the F555W-F814 colors, this filter combination is fully independent of the F343N-F555W colors.
It does have a slightly narrower color baseline than F555W-F814W and the predicted color difference between $Y=0.25$ and $Y=0.40$ RGB stars is reduced to $-0.028$ mag at $M_\mathrm{F606W}=+2$ (the $Y=0.25$ vs.\ $Y=0.33$ difference is $-0.015$ mag). Nevertheless, this should still be easily detectable.
We matched our WFC3 data with the ACSGCS photometry and defined a $\Delta$(F606W-F814W) index in the same way as for the WFC3 observations.  Figure~\ref{fig:acshist} shows the distributions of $\Delta$(F606W-F814W) colors for the group A, B, and C stars. 
We find $\langle \Delta(\mathrm{F606W}-\mathrm{F814W}) \rangle_A = -0.013$ mag, $\langle \Delta(\mathrm{F606W}-\mathrm{F814W}) \rangle_B = -0.009$ mag, and $\langle \Delta(\mathrm{F606W}-\mathrm{F814W}) \rangle_C = -0.012$ mag.
Note that the group A and C stars now have very similar $\mathrm{F606W}-\mathrm{F814W}$ colors, and the group B stars differ by only 0.004 mag from the group A stars. Formally, this corresponds to a difference in the He abundance of $\Delta Y \sim 0.021$, but the statistical significance of any differences between the color distributions of the three groups is small: a K-S test yields $P_\mathrm{AB}=0.05$, $P_\mathrm{AC}=0.60$, and $P_\mathrm{BC}=0.15$, respectively. 

From these comparisons, we conclude that a conservative upper limit on the He abundance variations in M15 is $\Delta Y\lesssim0.03$, which corresponds to a mass difference of $\lesssim0.04 \, M_\odot$ on the RGB for coeval populations.  As recently emphasized by \citet{Dotter2015}, absolute constraints on He abundance variations from CMD analyses remain uncertain. However, even with a generous allowance for model uncertainties, it seems difficult to accommodate the large variations in He abundance that are necessary in order for dynamical effects to produce the observed radial distributions of the sub-populations in M15.

\section{Discussion}

The radial trends in the ratios of N-normal to N-enhanced stars found in the preceding sections are very surprising, given that all current scenarios for the origin of multiple stellar populations predict that the enriched population should be more centrally concentrated. While the exact radial distributions predicted by each of the models are clearly subject to considerable uncertainty, it is difficult to imagine formation scenarios in which the enriched stars preferentially \emph{avoid} the center, as appears to be the case in M15. 

In the absence of an obvious explanation related to the formation of the populations, we have considered the possibility that dynamical evolution is responsible. We have argued that mass segregation might produce trends similar to those observed in the inner regions, but only if the enriched stars are very strongly He-enhanced. At 13 Gyr, the difference between models for giants with normal ($Y=0.25$) and the most He-enhanced composition available ($Y=0.40$) is about $0.2 \, M_\odot$, which is barely sufficient --  our $N$-body simulations indicate that a mass difference of $\sim0.25\,M_\odot$ or more is required. Further out, any trends set up at formation would be preserved.  Unfortunately, the currently available constraints on the He abundances of the different populations do not appear to support this explanation. Instead, we estimate that the masses of stars on the lower RGB differ by less than $0.04 \, M_\odot$, effectively ruling out mass segregation via two-body relaxation as a viable explanation for the observed radial trends.

Differences in the masses of red giants could also be produced by age differences, but in order for the enriched giants to be pushed outwards by mass segregation, they would have to be \emph{older} than the primordial ones. Furthermore, the age difference would have to be extremely large; even a difference of 3 Gyr corresponds to a mass difference of only $0.05 \, M_\odot$ at the turn-off. Exploring this possibility fully would probably require a dedicated $N$-body simulation that explicitly takes the age differences into account, but large age differences appear to be ruled out by the narrow subgiant branch in M15.

\subsection{Other constraints on He abundance}

It should be kept in mind that the relations between He abundance and the broad-band colors on the RGB remain to be verified observationally, although they appear to be a relatively solid prediction of the models. \citet{DiCriscienzo2011} found color differences between He-normal and He-enhanced RGB stars similar to those predicted by the \citet{Dotter2007} models, as did \citet{Beccari2013} based on BaSTI isochrones \citep{Pietrinferni2004}. From a self-consistent modeling of isochrones with enhanced He ($Y=+0.35$) and modified light element abundances, \citet{Salaris2006} found a difference in $V-I$ of $\sim0.03$ mag with respect to He-normal stars on the lower part of the RGB, again similar to the offsets in Fig.~\ref{fig:hecmp}. 

It would certainly seem worthwhile to obtain deep, high precision photometry of the main sequence in M15 in order to obtain better constraints on any variation in He abundances. 
In the meantime, another indicator worth exploring is the horizontal branch (HB) morphology. The suggestion that HB morphology and He abundance of GCs may be related is as old as the second parameter problem itself \citep{Sandage1967,VandenBergh1967,Kraft1979,Freeman1981}. M15 has a complex, bimodal HB morphology with an extended blue tail that has proven difficult to model in detail \citep{Buonanno1985a,Durrell1993,Moehler1995}. From the HB morphology and the RR Lyrae period distribution, \citet{DAntona2008} inferred a second-generation fraction of 80\% and a moderate He enhancement ($Y=0.26-0.30$). 
Using the isochrones of \citet{Dotter2007}, an $Y=0.30$ RGB star with an age of 13 Gyr has a mass of about 0.72 $M_\odot$, which is a difference of $-0.07 \, M_\odot$ with respect to the standard ($Y=0.25$) composition. \citet{Jang2014} suggested that the stars in the blue tail of the M15 HB belong to a population with $Y=0.33$ that accounts for 42\% of the stars in the cluster, but their models assume a 1 Gyr age difference between the first and subsequent generations of stars. In this case, the mass of a He-rich RGB star would be 0.69 M$_\odot$. From Figure~\ref{fig:m15seg}, even a $\sim0.1 \, M_\odot$ mass difference between the different populations seems insufficient to explain the observed radial trends as an effect of mass segregation.

While recent work has focused on He, it is well known that there are other parameters, such as the abundances of CNO, that can affect HB morphology \citep{Hartwick1972,Dorman1991,Dorman1992,Salaris2006,Milone2014}. Indeed,  \citet{DAntona2008} found that they were not able to get a satisfactory fit to the M15 HB by varying only He.
Data that allow comparison of He abundances derived from the HB and other methods are only available in a few cases. In 47 Tuc, \citet{DAntona2008} found a second-generation fraction of $\sim25$\% and $Y=0.27$--0.32 ($\Delta Y=0.02$--0.07) from the HB, whereas \citet{Milone2012a} found  $\Delta Y\sim 0.015$ and a second-generation fraction of $\sim70$\% from analysis of the full CMD, thus favoring a He enhancement toward the lower end of the range indicated by the HB analysis. Similarly, \citet{DiCriscienzo2010} found that the HB morphology of 47 Tuc and spread in the luminosity of stars on the sub-giant branch could be explained by a small but real He abundance spread of $\Delta Y = 0.02$, in agreement with the analysis of \citet{Milone2012a}, combined with an enhancement of the C$+$N$+$O sum in a fraction of the He-enhanced stars.
In NGC~6397, \citet{DAntona2008} suggested a negligible first generation fraction and $Y=0.28$ ($\Delta Y=0.04$) for the second generation, whereas \citet{Milone2012c} found that 70\% of the stars belong to an enriched population with $\Delta Y \sim 0.01$ ($Y\sim0.26$). 
Finally, in the case of NGC~2808, \citet{Dalessandro2011} found that the HB morphology could mostly be well matched by a model that incorporates the constraints on He abundance from the three distinct main sequences in this cluster, although this model had some difficulty reproducing the hottest ``blue hook'' stars on the HB.
These comparisons underline the considerable uncertainties involved in inferring He abundances from CMD analyses, and from the HB in particular, but also show that the extreme degrees of He enhancement in clusters like $\omega$ Cen and NGC~2808 are far from universal.

\subsection{Mixing of sub-populations}
\label{sec:mixing}

Any initial segregation of equal-mass sub-populations within a GC is expected to be eventually erased by mixing due to two-body relaxation \citep{Decressin2008}. However, this is a relatively slow process; the $N$-body simulations of \citet{Vesperini2013} indicate that any initial differences in the half-mass radii of different populations should remain detectable at least until the cluster is 10 half-mass relaxation times old (here referring to the current half-mass relaxation time). The current half-mass relaxation time of M15 is about $10^9$ years \citep{Djorgovski1993} so it does not seem unreasonable that the relatively moderate variations in the number ratios (by a factor of 2--3) observed in the central regions of M15 are still preserved.

\begin{figure}
\includegraphics[width=\columnwidth]{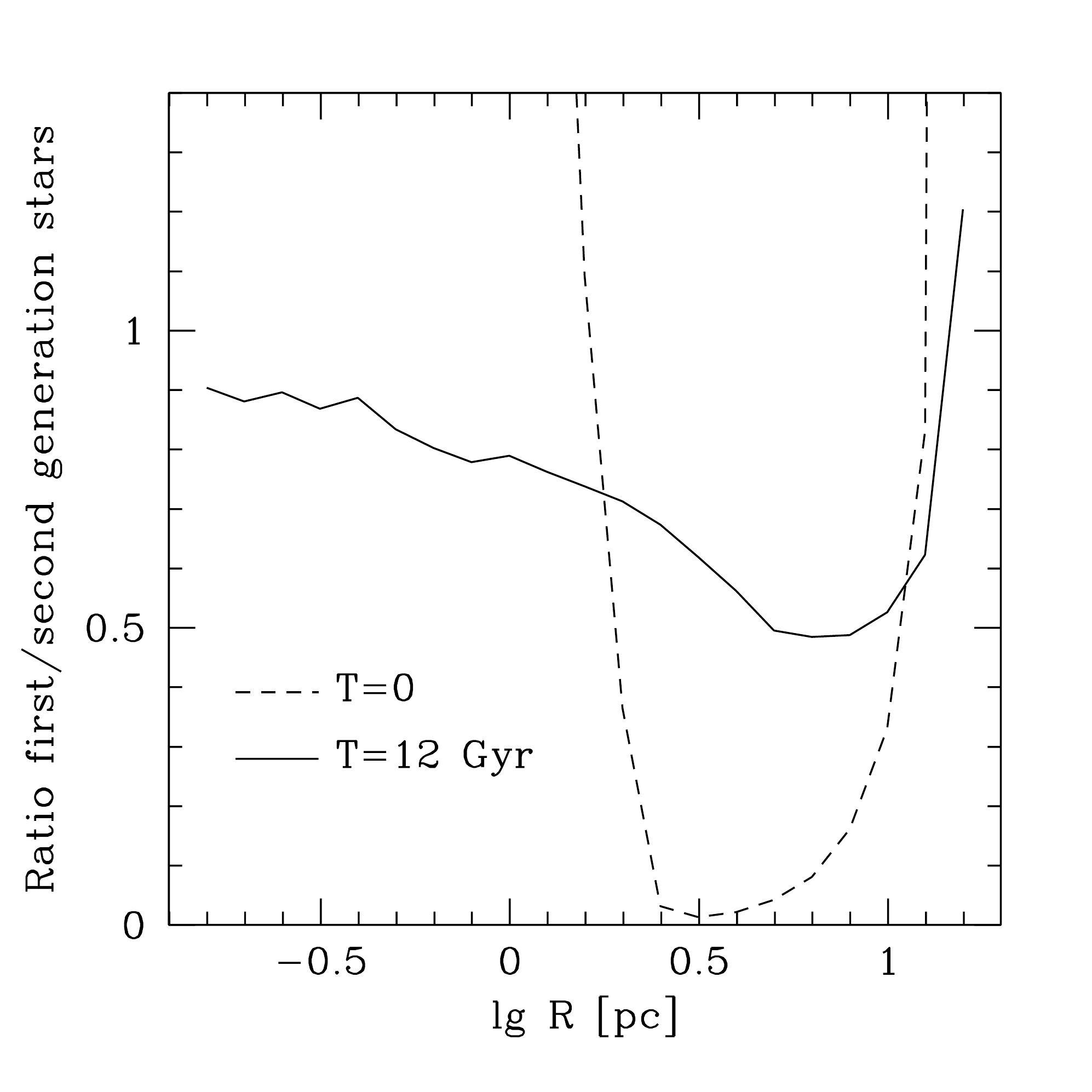}
\caption{\label{fig:relfrac}Ratio of primordial to enriched stars as a function of projected distance for the best-fitting $N$-body model to M15 as described in Sect.~\ref{sec:nbody} and \ref{sec:mixing}. For this plot we assumed that the stellar evolution of both stellar populations is the same. All stars with initially low and high orbital energies were assumed to be primordial stars while stars with intermediate orbital energies were assumed to be enriched stars. Despite orbital mixing due to two-body relaxation, a significant difference between the two populations is preserved until the end of the simulation at T=11.5 Gyr. }
\end{figure}

In order to check how quickly initial segregation is erased by two-body relaxation if the two sub-populations have equal mass, we again used the best-fitting model from the grid of \citet{McNamara2012} and assumed that all stars with initial orbital energies between $-1.0 \le E \le -0.2$ (in $N$-body units) are enriched stars while all stars with lower and higher energies are primordial stars. Figure~\ref{fig:relfrac} shows the projected number ratio of $M>0.6 \, M_\odot$ primordial to enriched stars that corresponds to this choice at the start of the simulations and after 12 Gyr of evolution. Due to two-body relaxation and resulting orbital mixing, the initial segregation between the two components is strongly weakened but not completely erased and the number ratio between the populations agrees qualitatively with what we find for M15 as shown in Fig.~\ref{fig:m15rmrg}.  We find similar amounts of remaining segregation for other choices of initial segregation, which demonstrates that the signal that we see is not due to one particular choice of initial segregation. The presence of a segregation of different stellar generations in M15 is therefore not in conflict with the idea that the populations have the same mass. 

\subsection{Are there other cases like M15?}

While the enriched population has generally been found to be more centrally concentrated than the primordial population within GCs, the majority of the studies to date have not looked at the central regions of clusters, largely due to resolution effects.  For example, \citet{Carretta2010}, \citet{Lardo2010}, \citet{Kravtsov2011}, Johnson \& Pilachowski (2012), and  \citet{Beccari2013} have all used ground based observations (photometry and/or spectroscopy) to study the relative spatial distributions of the populations, and have concluded that the enriched population is significantly more centrally concentrated than the primordial one.  The above studies have been focused outside the central $\sim1\arcmin$ for their respective clusters, typically corresponding to $1-2$~pc.  In Paper~I we found hints of the enriched population being more centrally concentrated than the primordial one in the four metal-poor GCs in the Fornax dSph, but again this was restricted to radii outside $1/2-1$ half-light radii.
As seen in our analysis of M15 \citep[also studied by][]{Lardo2010}, outside this radius the enriched population is indeed more centrally concentrated, however inside this radius a reversal occurs.  Hence, it is possible that such a reversal in the enriched/primordial population ratios in the central regions is a relatively common feature, and has  gone undetected due to resolution constraints.

A few HST studies have found that the enriched population does remain more centrally concentrated even in the core of the cluster.  \citet{Bellini2009} used HST imaging to study the central regions of $\omega$ Cen, and found that the enriched population does (slightly) dominate in the inner two core radii, but outside this radius, the enriched population is significantly more centrally concentrated that the primordial one.  \citet{Milone2012a} found that the enriched/primordial ratio increases toward the center in the massive GC 47 Tuc, although they have only two bins within the half-mass radius.

Finally, we note that other cases of more centrally concentrated primordial populations may already have been found. In NGC~2419, \citet{Beccari2013} found that giants with blue $u-V$ colors are more centrally concentrated than those with redder $u-V$ colors, i.e., the $u-V$ colors in NGC~2419 show the same behavior as the F343N--F555W colors in M15. Beccari et al.\ attributed the color differences to He abundance variations (so the blue stars would correspond to the enriched population), but also found that stars with anomalous (i.e., depleted) Mg abundances tended to have \emph{redder} than average $u-V$ colors. It would seem, therefore, that an alternative interpretation of their observations is that the giants with blue $u-V$ colors are, in fact, stars with normal (primordial) composition, and stars with red $u-V$ colors are enriched stars. NGC~2419 would then be similar to M15 in having a more centrally concentrated primordial population. However, it should be noted that the behavior of Mg in NGC~2419 is somewhat unusual, with $\mathrm{[Mg/Fe]}$ reaching very low values in some stars, and no obvious (anti-)correlation between $\mathrm{[Na/Fe]}$ and $\mathrm{[Mg/Fe]}$. NGC~2419 also displays other peculiar characteristics, including a large variation in $\mathrm{[K/Fe]}$  \citep{Cohen2012,Ventura2012,Beccari2013}.
Although the enriched stars in NGC~2419 might be more He-enhanced than those in M15, NGC~2419 is among the most extended GCs in the Milky Way and has very long central and half-mass relaxation times \citep[$t_{rc}\sim10.5$ Gyr and $t_{rh}\sim19$ Gyr;][]{Djorgovski1993}. It shows no evidence for significant mass-segregation \citep{Dalessandro2008,Baumgardt2009,Bellazzini2012}, and it might therefore also be difficult to explain differences in the radial distributions of sub-populations within NGC~2419 dynamically even if the populations have significantly different He abundances.

A recent ground-based study by Alonso-Garc\'ia et al. (in prep.), using Str\"omgren photometry of NGC 288, also found that the primordial population is more centrally concentrated than the enriched population, from $0\farcm5-7\arcmin$ ($\sim1.3-18$~pc) from the cluster core.

These cases, along with our results for M15, show that GCs display a wide variety of behaviors in the relative distributions of the enriched and primordial stars, in apparent contradiction with the standard scenarios for the origin of multiple populations within GCs.

\section{Summary and conclusions}

We have combined HST/WFC3 F343,F555W,F814W and SDSS $u,g$ observations to study the radial distributions of red giants with N-normal and N-enhanced composition over a radial range of 4$\arcsec$--600$\arcsec$ ($\sim0.06-10$ half-light radii) in the globular cluster M15. Our findings are as follows:

\begin{itemize}
\item The spread in the F343N-F555W colors of RGB stars of a given magnitude is far greater than the observational errors and implies a variation in [N/Fe] of about 2 dex, which is consistent with previous spectroscopic results.
\item Dividing the stars into three groups with ``primordial'' (i.e., similar to halo field stars), intermediate and strongly enriched nitrogen abundances (group A, B, and C), we find that the group A stars are the most centrally concentrated within the WFC3 field of view and the group C stars the least centrally concentrated. The difference is highly significant and contrary to the expectations from current scenarios for GC formation, which predict that the stars with primordial composition should be the least centrally concentrated. 
The group B stars have a degree of central concentration intermediate between the A and C groups, but the difference between the B and C stars is less significant.
\item When including the SDSS photometry, we find that the trend reverses in the outer parts of the cluster where the $N_A/N_{B+C}$ (primordial/enriched) ratio again increases. The fraction of primordial stars has a minimum near 1$\arcmin$, coinciding roughly with the half-light radius of the cluster.
\item $N$-body simulations indicate that a difference in mass of about 0.25 $M_\odot$ between the primordial and the most enriched giants could produce the observed radial trends due to two-body relaxation driven mass segregation. 
Such a mass difference could arise if the N-enriched stars also have a strongly He-enhanced composition ($Y\ga0.40$).
\item  However, the small differences in optical colors on the lower RGB suggest that there are \emph{no} large differences in the He abundances of primordial and N-enriched stars ($\Delta Y \la 0.03$), with a corresponding mass difference of less than $0.04 \, M_\odot$ if the stars have the same ages.
\end{itemize}

We are thus left with no convincing explanation for the observed radial distributions of different stellar populations in M15. 
We find no evidence that variations in the foreground reddening might cause the observed trends; a differential reddening correction based on the F606W-F814W colors of main sequence stars actually increases the significance of the trends. However,  small color variations across the field might also be caused by instrumental effects. Data in more passbands would be required in order to quantify these effects better.

If the overall trends found here, including the lack of a significant difference in He abundance, are confirmed (for example by deep, multi-passband photometry of the main sequence), then the differences in the central regions are unlikely to be of a dynamical origin and, presumably, must reflect the conditions at the time of formation. We have shown that such differences could be preserved until the present epoch, due to the relatively slow nature of orbital mixing. This would represent a challenge to all current scenarios for the formation of GCs.

\acknowledgements
We thank Henny Lamers for discussion about relaxation time scales. The anonymous referee provided a helpful and detailed report that led to significant improvement of the paper.
JB and JS acknowledge support for HST Program number GO-13295 from NASA through grants HST-GO-13295.02 and HST-GO-13295.03 from the Space Telescope Science Institute, which is operated by the Association of Universities for Research in Astronomy, Incorporated, under NASA contract NAS5-26555. JB also acknowledges HST grant HST-GO-13048.02 and NSF grant AST-1109878.
Funding for the Stellar Astrophysics Centre is provided by The Danish National Research Foundation. The research is supported by the ASTERISK project (ASTERoseismic Investigations with SONG and Kepler) funded by the European Research Council (Grant agreement no.: 267864).
This research has made use of the NASA/IPAC Extragalactic Database (NED), which is operated by the Jet Propulsion Laboratory, California Institute of Technology, under contract with the National Aeronautics and Space Administration.

Facilities: \facility{HST(WFC3)}, \facility{HST(ACS)}, \facility{SDSS}

\bibliographystyle{apj}
\bibliography{libmen.bib}

\end{document}